\documentclass[a4paper,11pt]{article}
\pdfoutput=1 

\usepackage{jcappub} 
\usepackage{multirow}

\usepackage[T1]{fontenc} 
\usepackage[modulo]{lineno}
\usepackage{amsmath}

\title{\boldmath Online triggers for supernova and pre-supernova neutrino detection with cryogenic detectors}

\author[a,1]{P. Eller}\note{Corresponding authors.}
\author[b,1]{N. Ferreiro Iachellini}
\author[a,c,1]{L. Pattavina}
\author[b,1]{and L. Shtembari}

\affiliation[a]{Physik-Department and Excellence Cluster Origins, Technische Universit{\"a}t M{\"u}nchen \\ James-Franck-Stra{\ss}e 1, DE-85747 Garching, Germany}
\affiliation[b]{ Max-Planck-Institut f{\"u}r Physik, \\ F{\"o}hringer Ring 6, DE-80805 M{\"u}nchen, Germany}
\affiliation[c]{INFN Laboratori Nazionali del Gran Sasso, \\ Via G. Acitelli 22, I-67100 Assergi, Italy}

\emailAdd{philipp.eller@tum.de}
\emailAdd{ferreiro@mpp.mpg.de}
\emailAdd{luca.pattavina@lngs.infn.it}
\emailAdd{lolian@mpp.mpg.de}

\abstract{Supernovae (SNe) are among the most energetic events in the universe still far from being fully understood. An early and prompt detection of neutrinos is a one-time opportunity for the realization of the first multi-messenger observation of these events. In this work, we present the prospects of detecting neutrinos produced before (pre-SN) and during a SN while running an advanced cryogenic detector. The recent advancements of the cryogenic detector technique and the discovery of coherent elastic neutrino-nucleus scattering offer a wealth of opportunities in neutrino detection. The combination of the excellent energy resolution of this experimental technique, with the high cross section of this detection channel and its equal sensitivity to all neutrino flavors enables the realization of highly sensitive cm-scale neutrino telescopes, as the newly proposed RES-NOVA experiment. We present a detailed study on the detection promptness of pre-SN and SN neutrino signals, with direct comparisons among different classes of test statistics. While the well-established Poisson test offers in general best performance under optimal conditions, the non-parametric Recursive Product of Spacing statistical test (RPS) is more robust and ideal for triggering astrophysical neutrino signals with no specific prior knowledge. Based on our statistical tests the RES-NOVA experiment is able to identify SN neutrino signals at a 15~kpc distance with 95\% of success rate, and pre-SN signal as far as 450~pc with a pre-warn time of the order of 10~s. These results demonstrate the potential of RPS for the identification of neutrino signals and the physics reach of the RES-NOVA experiment. }

\begin{document}
\maketitle

\section{Introduction}
\label{sec:intro}
The era of neutrino astronomy started in 1987 when neutrinos from the first extra-galactic source were detected~\cite{Kamiokande1987}. This was the renowned SN1987A which produced in total about 25 events in different detectors around the world~\cite{Review1987neutrino}. Only 30 years later, a second source was identified TXS-0506+056~\cite{IceCube:2018dnn, TXS}. This was a Blazar, and it marked the birth of multi-messenger astronomy, thanks to the detection of its electromagnetic counterpart.
Having the possibility to predict when and where the next high-energy cosmic event will take place, it will be a major breakthrough, because it may lead to the \textit{Holy Grail} of multi-messenger astronomy: the simultaneous detection of neutrinos, electromagnetic radiation and gravitational waves.

Supernovae (SNe), which are among the most energetic events in the Universe, are interconnected with many aspects of astroparticle physics and astronomy. In fact, when massive stars ($>8~M_\odot$) become SNe almost their entire binding energy is released as neutrinos~\cite{Baade254}. These high intensity neutrino fluxes are one of the seed for the nucleosynthesis of heavy elements, at the same time the remnants of the explosions can act as accelerator for Ultra-High-Energy cosmic rays, but they can also trigger stellar formation~\cite{Mirizzi:2015eza}.
The detection of neutrinos is key in this framework for two reasons: first, they are direct probes of the central engine of the SN and they are thought to be responsible for the stellar explosion; second, they can provide an early alert of the forthcoming event. Actually, neutrinos are emitted prior to the gravitational collapse determining the SN event, anticipating it by minutes or even days~\cite{Nakamura_2016}. These ``pre-Supernova'' neutrinos are produced at the late stage of Si burning and give insight into processes in the deep interior of massive stars prior to their deaths~\cite{Odrzywolek:2003vn}. When these neutrinos are observed, they can anticipate the explosion as early as a few hours or days~\cite{Asakura:2015bga}, enabling the preparation of all available detection techniques for all the emission components of the event, namely gravitational waves, electromagnetic components, and neutrinos. For this reason these are usually defined as pre-Supernova (pre-SN) neutrinos. 

Currently, pre-SN neutrinos were never observed. The challenge in detecting them is two-fold: the first reason concerns the low rate of SN events, especially the galactic ones. Second, compared to conventional SN neutrino emission, pre-SN neutrinos have lower average energies ($< 5$~MeV, while for a SN event $\gg 10$~MeV). Furthermore, flux intensities are many orders of magnitude lower, even when considering nearby events ($\ll$1~kpc). However, the chances of observing these weak signals are still not null because the galactic SN rate is not uniform throughout the Milky Way~\cite{Adams:2013ana}. Especially if we look at all the 6 historically recorded galactic SNe, they all occurred at distances $<$ 3~kpc from us~\cite{The:2006iu}. This represents about 20\% of the galactic rate, although this region represents only 4\% of the area of the galactic disk. This demonstrates that the Earth is in a very active region of the Milky Way~\cite{Firestone_2014}, and it is likely that the next SN event will occur in our vicinity. In addition, the latest technological advancements and the upgrade of detectors currently operating with advanced signal discrimination capabilities have also enhanced their sensitivity to such weak signals. In this framework, the recent Super-Kamiokande upgrade~\cite{Simpson:2019xwo} with Gd will enable a more efficient discrimination of inverse-beta decay events~\cite{Simpson:2019xwo} enhancing its sensitivity to the pre-SN signal~\cite{Super-Kamiokande_PreSN}. Moreover, the recent discovery of Coherent Elastic neutrino-Nucleus scattering (CE$\nu$NS)~\cite{Freedman:1973yd} has broadened the perspectives of observing this signal using a neutral current channel. In fact, as noted in~\cite{Raj:2019wpy}, CE$\nu$NS is an ideal channel for the detection of pre-SN neutrinos given: its high cross section, $\gg 10^3$ times higher than the inverse beta decay (IBD) one, equal sensitivity to all neutrino flavors and lack of a kinematic energy threshold, so that also neutrinos with energies lower than 1.8~MeV can be detected~\cite{Drukier:1983gj}.

Among the different experimental techniques which can exploit CE$\nu$NS for studying the properties of neutrino sources, cryogenic calorimetric detectors are the most promising ones. In fact, they demonstrated to achieve ultra-low energy thresholds and precise energy reconstruction~\cite{DM_cryo} (no energy quenching), which is relevant for the detection of low energy nuclear recoils induced by neutrino scattering via CE$\nu$NS. In addition, in recent years the scalability of this technology to large volume arrays was also demonstrated~\cite{CUORE_Nature}, as well as the flexibility in operating different target materials~\cite{Pattavina:2018nhk,Casali:2013zzr, Beeman:2011kv, Cardani:2013dia, Beeman:2012wz, Pattavina:2015jxe, Artusa:2016mat}.

In this manuscript we will review the methods for detecting neutrinos from the late Si burning stage of stars with cryogenic detectors. The main focus of this work will be a study of the expected neutrino signal induced in different target materials (i.e., PbWO$_4$, CaWO$_4$ and Ge) with demonstrated promising features as cryogenic particle detectors. We will also investigate advanced statistical trigger methods for a prompt online detection of CC-SN neutrinos. Finally,  we will show the potential of the previously discussed online triggers for the specific case of pre-SN neutrinos, with the aim of providing multi-messenger alert to the scientific community before the SN event occurrence.

\section{Pre- and Supernova neutrinos}
\label{sec:preSN}

\begin{figure}[t]
\centering
\includegraphics[width=0.9\textwidth]{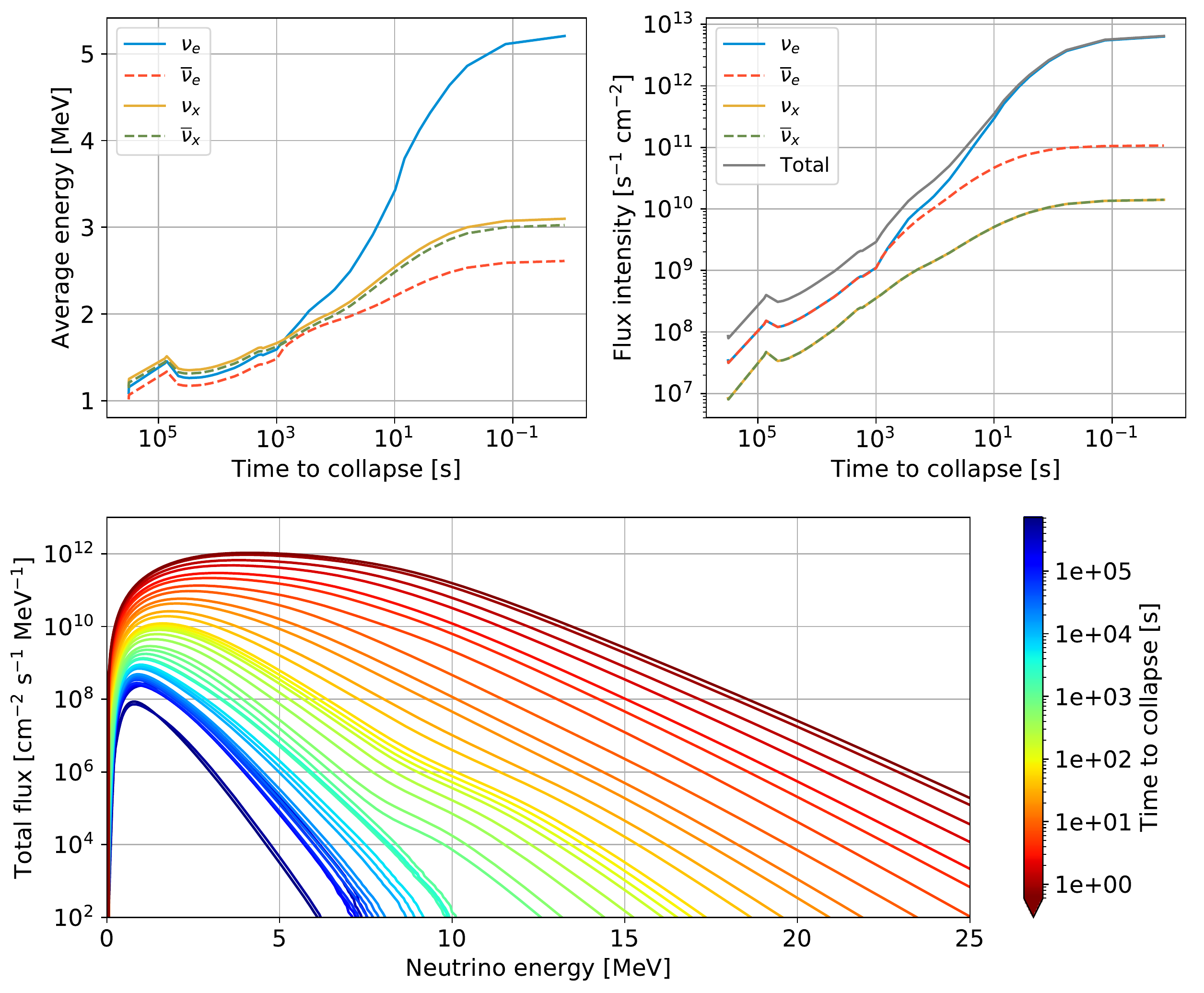}
\caption{Characteristics of the pre-Supernova neutrino emission from a progenitor star with $15~M_\odot$ mass placed at 160~pc (same distance as Betelgeuse~\cite{Joyce_2020}). The model considered is~\cite{Kato:2017ehj}.}
\label{fig:preFlux} 
\end{figure}

Two types of emissions drive the production of pre-SN neutrinos: thermal pair processes and nuclear weak processes~\cite{Kato:2017ehj}. The following reactions fall within the first class of emissions: $e^-$-$e^+$ annihilation, plasmon decay, neutrino bremsstrahlung and photo-processes. Depending on the progenitor star's mass, namely density and temperature, the rate of the different processes may vary. For massive stars with mass $>8~M_\odot$, the main reaction channel is pair annihilation~\cite{Odrzywolek:2003vn}. All these processes produce (anti-)neutrino of all flavors ($\nu_x$/$\bar\nu_{x}$, for $x=\mu,\tau$). 
The second type of neutrino emission, which becomes dominant just before the core-collapse~\cite{Patton:2015sqt,Odrzywolek:2009wa} (about 400~s before the core-collapse), are nuclear weak processes such as: electron-capture, $\beta^-$/$\beta^+$-decay and positron captures. These are the main source of $\nu_e$/$\overline{\nu}_e$, which make the largest fraction of the pre-SN neutrino emission.
Clearly, the neutrino flux intensity and average energies increase for all flavors as the time of the collapse approaches. In fact, the core temperature and density increases yielding to neutrino of higher energies. The relevance of nuclear weak processes is visible in Fig.~\ref{fig:preFlux}, where at 400~s before the collapse when electron-captures become the dominant reaction. There, the average energy and flux intensity of $\nu_e$ increase at a higher pace than all the other flavors, leading to a pre-SN neutrino flux composed primarily by $\nu_e$.
Some structures are also visible in the time profile of the pre-SN neutrino emissions, like at 10$^5$~s which corresponds to the transition between the O and the Si burning stages. A detailed discussion of all the processes that steer the evolution of neutrinos can be found in~\cite{Kato:2017ehj,Patton:2015sqt,Odrzywolek:2009wa}.

Pre-SN neutrino fluxes are expected to be weaker than the signal produced during a SN event. As shown in Fig.~\ref{fig:preFlux}, at about 100~s prior the core-collapse the total flux intensity is about 10$^{10}$~$\nu$/cm$^2$/s, while assuming the neutrino source at a distance of 160~pc. The same progenitor star would produce a neutrino flux with an intensity of 10$^{16}$~$\nu$/cm$^2$/s during the core-collapse of SNe (CC-SNe).
State-of-the-art detectors have masses of $\mathcal{O}(10~\textrm{kton})$~\cite{SNEWS:2020tbu} that enable the possible detection of these feeble neutrino signals. The golden channel for their detection is IBD, but this is still not the ideal channel given the limited cross-section ($\sigma_{IBD} \sim 10^{-41}$~cm$^2$ for 10~MeV neutrinos), and the kinematic threshold allows the detection of only the higher energy tail of the neutrino thermal distribution. Very few events are expected to be detected even when the source is in the vicinity of Earth, $\mathcal{O}(200~\textrm{pc})$.
CE$\nu$NS might offer the opportunity to overcome the limitation of the present technology, due to its high sensitivity~\cite{Raj:2019wpy} to the entire spectrum of pre-SN emission. Depending on the type of target material, a cross section as high as $10^{-38}$~cm$^2$ for 10~MeV neutrinos can be achieved.
A not exhaustive list of SN candidates~\cite{Nakamura_2016} that might enable the detection of pre-SN neutrinos contains: Antares (150~pc), Betelgeuse (160~pc), $\epsilon$-Pegasi (210~pc), $\pi$-Puppis (250~pc), $\sigma$-Canis Majoris (340~pc), NS Puppis (520~pc), CE Tauri (550~pc) and 3 Ceti (640~pc).

Similar arguments apply to the detection of CC-SN neutrinos, but the achievable sensitivities do not differ much between the conventional technologies and the one based on CE$\nu$NS. This is due to the increase of the average neutrino energies above the critical threshold of 1.8~MeV. However, CE$\nu$NS detectors offer the unique possibility of detecting with high statistics the $\nu_x$/$\bar\nu_{x}$ component.


\section{Pre-Supernova and Supernova neutrino detection with cryogenic detectors}
\label{sec:detector}

Cryogenic calorimeters are ideal devices for the detection of low energy nuclear recoils. Details on the working principle of cryogenic detectors can be found in the following references~\cite{Pirro:2017ecr, Kim:2021wae, MUNSTER2017387}. This technology was demonstrated to be effective, especially in direct Dark Matter (DM) experiments, where the expected signals are nuclear recoils with very low energies $\ll 1$~keV range~\cite{Appec_DM}, the same type of signature expected from CE$\nu$NS interactions.

The most relevant aspects for the study of CE$\nu$NS interactions from SN neutrinos concern the achievement of low-energy detector thresholds $\mathcal{O}(100~\textrm{eV})$ and the background level in the region of interest (RoI) $\mathcal{O}$($\ll$ 1~c/keV/kg/d).

\subsection{Signal expectation}
\label{sec:detector_signal}
For our sensitivity studies on the detection of the feeble pre-SN neutrino signal with cryogenic detectors, we consider three target materials: PbWO$_4$, CaWO$_4$ and Ge. Each of these compounds has shown to achieve excellent detector performance. In fact, leading limits in the DM-regular matter interaction parameter space are established with these materials. CaWO$_4$ and Ge crystals are operated as cryogenic detectors for the search of DM by the CRESST~\cite{Abdelhameed:2019hmk, CRESST:2019mle} and SuperCDMS/Edelweiss experiments~\cite{SuperCDMS:2022kse,Armengaud:2019kfj}, respectively. They achieved leading detector energy thresholds, as low as 30~eV~\cite{Abdelhameed:2019hmk} and low-background in the RoI, about 1~c/keV/kg/d~\cite{Abdelhameed:2019oxl}. PbWO$_4$ is also a very promising candidate for our studies as widely discussed in~\cite{Pattavina:2020cqc}. This crystal is expected to reach outstanding background level when produced from archaeological Pb~\cite{Pattavina:2019pxw}. The RES-NOVA experiment~\cite{RES-NOVA:2021gqp} is planning to operate these crystals, and interesting results have already been achieved in terms of the energy threshold and the background level~\cite{RES-NOVAgroupofinterest:2022pvc, Iachellini:2021rmh}.
All of these compounds cover a wide range of target masses, enabling exploration of a broad parameter space for neutrino-nucleus interactions.

\begin{figure}[t]
\centering
\includegraphics[width=0.95\textwidth]{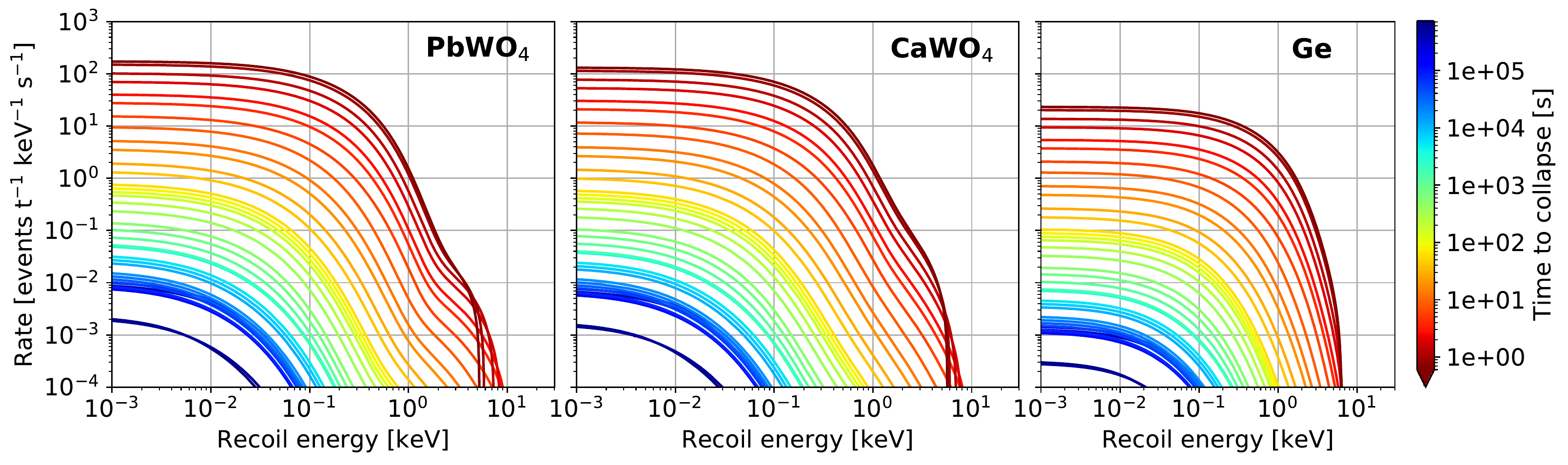}
\caption{Recoil energy spectra produced by the neutrino signals shown in Fig.~\ref{fig:preFlux}. The three considered target materials are: PbWO$_4$, CaWO$_4$ and Ge. The different recoil spectra colors are produced by pre-SN signal at different times anticipating the SN core-collapse. The SN is assumed to be  at 160~pc (same distance as Betelgeuse~\cite{Joyce_2020}). Data are taken from~\cite{Kato:2017ehj}.}
\label{fig:recoilpreFlux} 
\end{figure}

The detector responses to the pre-SN neutrino signals shown in Fig.~\ref{fig:preFlux} are represented in Fig.~\ref{fig:recoilpreFlux}. These are the recoil spectra produced by neutrino interactions via CE$\nu$NS. These are computed multiplying the number of target nuclei, $N_{target}$, for the total neutrino energy spectrum at time $t_{ttc}$ (time to collapse) $\phi(E_{\nu}, t_{ttc})$, and for the cross-section, $d\sigma_{CE\nu NS}/E_R$. These quantities are then integrated over the different neutrino energies :
\begin{equation}
\frac{d R}{d E_R} = N_{target} \int_{E^{min}_\nu} d E_{\nu} \; \phi(E_{\nu}, t_{ttc}) \;\frac{d \sigma_{CE\nu NS}}{d E_R}.
\end{equation}

The differential scattering cross section of CE$\nu$NS interactions is:
\begin{equation}
    \frac{d \sigma_{CE\nu NS}}{E_R} = \frac{G^2_F}{8\pi (\hbar c)^4} \left( \left( 4 \sin^2 \theta_W -1 \right) \cdot Z + N \right)^2 \cdot m_{N} \cdot\left( 2- \frac{E_R m_{N}}{E^2_{\nu}}\right) \left|f(q)\right|
    \label{eq:xs}
\end{equation}
where $\theta_W$ is the Weinberg angle,\footnote{$\left|4\sin^2 \theta_W -1\right| \sim 0.108$} $Z$ and $N$ are the proton and neutron numbers of the target nucleus, $m_{N}$ the mass of the target nucleus, $E_R$ the energy recoil of the target nucleus, $E_{\nu}$ the neutrino energy and $f(q)$ is the form factor. The latter term describes the coherency of the process as a function of the transferred momentum $q=\sqrt{2m_N E_R}$. The experimentally most relevant dependence of Eq.~\ref{eq:xs} is $N^2$, favoring the use of target material with large neutron number. In this context, Pb-based detectors have a great potential.

As discussed also in Sec.~\ref{sec:preSN}, the intensity of the neutrino flux increases as the time to the collapse approaches. This is shown in Fig.~\ref{fig:recoilpreFlux} by the higher interaction rate as the collapse approaches. In addition, the higher-energy end of the recoil spectra shifts to higher energies, due to the increasing average neutrino energies. In the same figure, the characteristic features of the CE$\nu$NS process are also recognizable:
\begin{itemize}
    \item Target nuclei with higher neutron number lead to higher interactions rates: $R_{PbWO_4} > R_{CaWO_4} > R_{Ge}$.
    \item Lighter target nuclei feature higher energy recoils: $Er^{max}_{Ge} > Er^{max}_{CaWO_4} > Er^{max}_{PbWO_4}$.
\end{itemize}

\begin{figure}[t]
\centering
\includegraphics[width=0.8\textwidth]{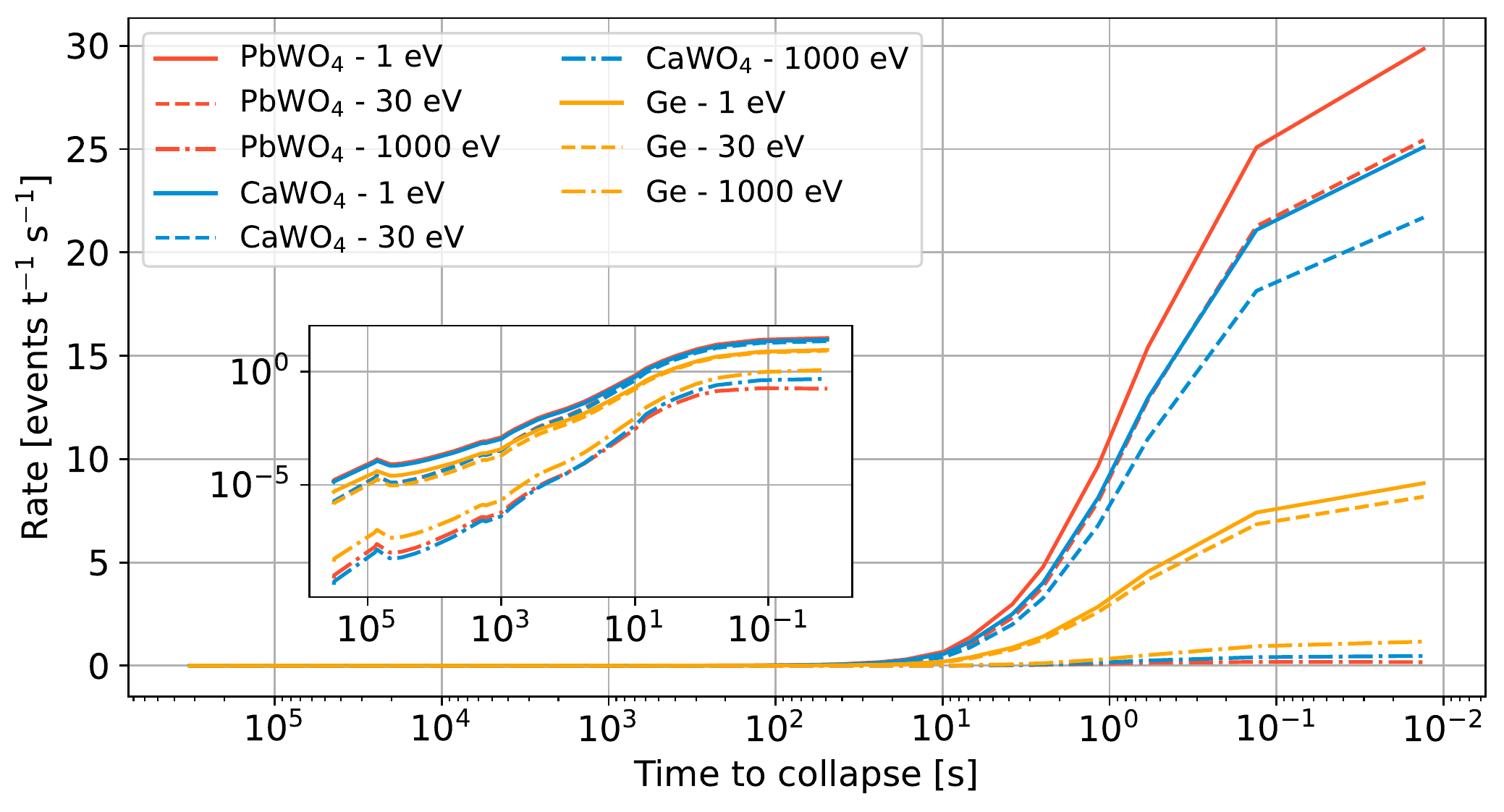}
\caption{Instantaneous interaction rates as a function of the time prior to the SN core-collapse. These are evaluated for three different target materials and different energy thresholds. In the inset, a zoom-in of the times with lower rates is shown.}
\label{fig:recoiltimepreFlux} 
\end{figure}

Experiments that wish to detect pre-SN neutrinos at earlier times need to operate large volume detectors, at the ton scale, and they need to operate in ultra-low-background conditions and low-energy threshold. In Fig.~\ref{fig:recoiltimepreFlux}, the instantaneous interaction rate for different target materials and different thresholds is shown. This is calculated by integrating each recoil spectra of Fig.~\ref{fig:recoilpreFlux} between the detector energy threshold (300~eV, 30 eV and 1 eV) and 50 keV. To anticipate a SN event through the detection of pre-SN neutrinos background counting rate in the RoI of $< 10^{-2}$~cts/ton/keV/s are needed, as well as an energy threshold $\ll 1$~keV.

In the following, we will consider only PbWO$_4$ crystals as target material, given that this ensures the highest neutrino counting rate among the candidate compounds, and thus it enables to provide longer pre-warn time for SN events. The RES-NOVA experiments, which aims at deploying an advanced neutrino telescope, is planning to use an array of PbWO$_4$ crystals produced from archaeological Pb, with a total active volume of only (60~cm)$^3$ (equivalent to 1.8~tons). RES-NOVA is planning to achieve a background level of about 10$^{-3}$~cts/ton/keV/s in RoI for the detection of pre- and CC-SN neutrinos, while operating in anti-coincidence mode~\cite{RES-NOVA:2021gqp}. 

The RoI, that will be considered in the following, for the detection of pre-SN neutrinos with PbWO$_4$ detectors lies between 1~eV and 4~keV. We are assuming an optimistic case of achieving 1~eV of energy threshold, for defining a benchmark \textit{best case scenario}, that allows the detection of pre-SN neutinos with a pre-warn time of $> \mathcal{O}(10\,\textrm{s})$.  For the RES-NOVA experiment the expected background rate in this RoI is 0.018~cts$\cdot$s$^{-1}$ (2.5$\cdot$10$^{-3}$~cts/ton/keV/s $\times$ 1.8~tons $\times$ 4~keV). This is the value that will be adopted for our sensitivity studies on pre-SN neutrinos.

The expected signal produced from a CC-SN in a cryogenic PbWO$_4$ detector was already presented in other works. For the sake of simplicity we invite the reader to look at Fig.~5 of ref.~\cite{Pattavina:2020cqc}. There the RoI extends from the detector energy threshold, assumed at 1~keV, up to 40~keV. In this case, we decide to consider a more realistic energy threshold, as a very optimistic one (e.g. 1~eV) will not significantly improve the expected neutrino signal amplitude. For CC-SN neutrino events in the [1,40]~keV RoI we consider a background rate of 0.18~cts$\cdot$s$^{-1}$ (2.5$\cdot$10$^{-3}$~cts/ton/keV/s $\times$ 1.8~tons $\times$ 40~keV).



\subsection{Background considerations}
\label{sec:bkg_estimation}


In the following sections we will develop a trigger for the live detection of signals from CC-SN, failed CC-SN and pre-SN neutrinos. The background clearly plays crucial role in the identification of a positive signal. It has to be pointed out that, in a real setup, especially when dealing with low-background experiments, the radioactive backgrounds are difficult to assess to the very last count and do show time dependence. Uncertainties on these spoil the application of simple Poisson statistics for the determination of expected rates and the confidence on them. Furthermore, not all backgrounds can be attributed to radiogenic origin. An example of this comes from the CRESST experiment, where it is observed that there are time periods (minutes of duration) where the trigger rates are substantially higher than the standard operating conditions~\cite{Abdelhameed:2019hmk, CRESST:2015txj}, due to instabilities of the cryogenic system. The cryogenic technique that we investigate in this work is well optimized for DM searches, where all these issues can be addressed in off-line analysis \cite{Reindl:2016yhs}. The application of such technology for the live and prompt detection of transient signals of astrophysical origin must not rely on human analysis and therefore must be fully automated and robust against such kinds of disturbances.

For what concerns the radioactive background, the best possible estimation for the background rate $r_{bkg}$ is its direct measurement and monitoring once the experiment is set in operation: the background rate can be extracted from a selection of collected data using Monte-Carlo methods and once an estimate is available it can be used to define the parameters of the analysis as we discuss in Sec.~\ref{sec:statistical_analysis}. The situation is more complicated when dealing with the other sources and, because of that, in the following we will present other test statistics beyond Poisson counting.

\section{Early identification of neutrino signals}
The next galactic SN will bring information about the physics processes that cannot be studied in any terrestrial experiment, and the elusive rate of such an event makes this information extremely valuable. For the first time in history, technologies to detect neutrinos, gravitational waves, and electromagnetic radiation from SN events are in place. It is of uttermost importance to record all possible data in the best quality, and to do so, the SN event needs to be detected as early as possible. 

The Supernova Early Warning System (SNEWS) is an international group of neutrino sensitive experiments aiming at providing the astronomical community with an early alerts for SN events. Being able of combining the signal from experiments sited at different locations on the globe brings several advantages. Firstly, it allows to increase the detection sensitivity, especially for weak signals coming from distant SNe. Secondly, it suppresses the Poissonian background fluctuations combining signals from experiments located at different laboratories, so that backgrounds are uncorrelated.

A multi-messenger observing strategy is key to fully exploit the wealth of information carried away by neutrinos. The neutrino emission starts before the CC even begins, meaning that neutrinos can provide an early warning signal. Knowing when and possibly where to anticipate the signal dramatically improves the detection prospects. During the stellar CC the neutrino emission is accompanied by the emission of GWs.  As discussed in~\cite{Nakamura_2016}, the neutrino arrival time can also act as a SN trigger, increasing the sensitivity of GW experiments. In addition, an early detection of neutrinos, and possibly pre-SN neutrinos, can anticipate the electromagnetic burst by several minutes or days~\cite{Nakamura_2016}.

\subsection{Time response in cryogenic particle detectors}
\label{sec:time_resp}
In cryogenic detectors particle interactions cause a temperature increase in the sensitive volume that is measured by means of a thermometer. Best sensitivities at low energies are achieved with TES (Transition Edge Sensors), where a superconductor is kept in an intermediate state between the normal conducting and the superconducting regime and the excellent sensitivity is determined by the steepness of the resistance value as a function of the temperature~\cite{Pirro:2017ecr}. The temperature rise induces a measurable variation of the TES resistance, which is related to the energy deposited in the absorber.

The development of thermal signals in TESs~\cite{Franz} directly depends on the specific geometry and layout of the sensor. A thermal pulse of a few keV of energy on a hundreds gram absorber develops over a time scale of about 500~ms~\cite{CRESST2}. This parameter is particularly important when designing an online trigger, especially when detecting SN neutrinos. To achieve maximum sensitivity at trigger level, the detector datastream is fed through a non-causal matched filter~\cite{FerreiroIachellini:2019obk}. The non-causality of the filter requires a software processing in Fourier domain that, in order to avoid Gibbs distortion~\cite{Gibbs}, needs to process signals that are of duration twice as long at the signal itself. This extra-duration is, in fact, simply a set of extra digitized samples, half of them before the signal and half past the signal. This effectively introduces a delay of half duration of the original signal, and the time between the actual occurrence of the (single) particle interaction and its identification is, at minimum, of about 750~ms.
The time development of signals in cryogenic detetors is rather long when compared with other technologies (e.g. PMT in liquid scintillator). However, the high modularity of these detectors enable to reduce to a negligible level the detector dead-time caused by pile-up events, as demonstrated for the RES-NOVA detector in~\cite{RES-NOVA:2021gqp}. 



\section{Statistical tools and data processing}\label{sec:statistical_analysis}

\begin{figure}[h]
\centering
\includegraphics[width=0.8\textwidth]{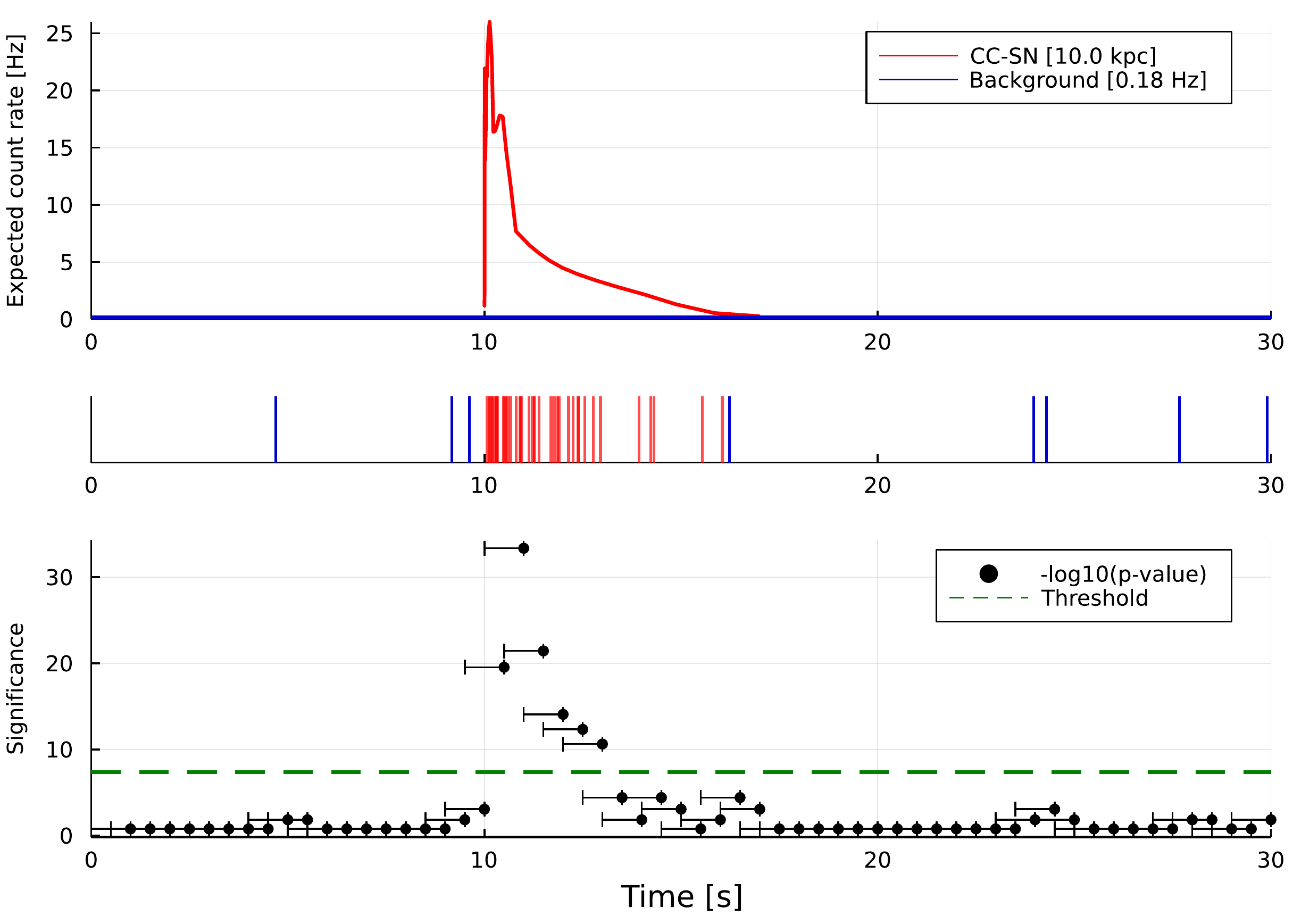}
\caption{Example of a CC-SN signal and Poisson analysis:
(top) Expected count rate for a SN at $t=10$\,s at distance of 10\,kpc, plotted together with a constant background rate of 0.18\,Hz.
(middle) Random realization of counts as seen in the detector.
(bottom) Analysis using 50\% overlapping windows of 1\,s length. For each window the Poisson p-value is calculated and indicated with the blue dot for the window extending 1\,s into the past. The green line gives a threshold that would yield a FAR of 1 per 15 days.}
\label{fig:post_SN_dist} 
\end{figure}

This section discusses approaches for building a triggering system to detect transient events---such as SNe---based on the real-time data stream of a neutrino detector.
Detected signal events are interspersed with background events that, for now, we will consider to be distributed according to a fixed rate Poisson process, of which we know the true rate.
The top panel of Fig.~\ref{fig:post_SN_dist} shows an example of the expected count rate in a RES-NOVA like detector for neutrinos from a SN at a distance of 10~kpc, together with the uniform expectation of background events.
The middle panel show how an example data stream could look like, generated from random variates of the expected count rate.
We want to find a statistical method that can deliver a yes/no answer in near real time to decide whether there was a SN signal present in the data. The false alarm rate (FAR) (i.e. type i error rate)---as an external constraint to our system---can on average be no below 1 per week, as required for participation in SNEWS~\cite{SNEWS:2020tbu}.

A standard way to analyze such time series data is to use fixed or variable length windows in time, and reduce the task at hand to decide whether the events inside a window exceed the expected Poisson count from the background rate.
If the background rate is exactly known, and the windows are non-overlapping, the threshold for a given FAR can be directly calculated from the Poisson distribution.
Since such a configuration depends on the location of the time window edges, often overlapping windows are used. For 50\% overlapping windows, approximations for the calculation of the p-value can be used. Such overlapping time window Poisson tests represent the current state-of-the-art, as used, for example, in~\cite{Agafonova:2007hn}. Alternative approaches have been proposed, for example, with dynamic time windows~\cite{Lamoureux:2021bat}.
The bottom panel of Fig.~\ref{fig:post_SN_dist} shows the Poisson p-values for 1\,s, 50\% overlapping windows for our example, and a threshold value resulting in a FAR of 1 per 15 days.

This type of system has a few parameters, including the background rate $r_{bkg}$ that can be estimated from background-only data, and the window configuration. There is a balance between choosing the window size $\omega$ large enough so that most of the signal is contained within the window and at the same time small enough so that the signal events will not be washed out by an additional background contribution.
The refresh rate, i.e., how often is a window analyzed, is another parameter of choice. For a refresh rate of $1/\omega$ we have the configuration of non-overlapping windows, and $2/\omega$ would correspond to the 50\% overlap.
Since we want to have the freedom to explore more complicated window choices, and later also different statistical tests, we first introduce a simulation based method to calculate critical values for a desired FAR.

\subsection{Computation of critical values}

Given a custom test that operates at a certain refresh rate, we can calculate a test statistic value $TS$, which for example could be the Poisson p-value itself.
However, for overlapping windows this quantity $TS$ can no longer be interpreted as the p-value of the test, and its distribution is in general unknown. To make a statistical statement about $TS$ we need to know, or rather estimate, its cumulative distribution $F_{TS}$.

The estimation $F_{TS}$ can be obtained using simulated data, producing values of $TS$ for the detection of neutrinos with a predetermined rate of the background-only scenario.
The simulation cannot be done with independent simulations, since this would remove the important correlations of successive values of $TS$ in consecutive time windows. Therefore, we simulate an extended run of the experiment and collect the values of $TS$ in a serial fashion, at least for time scales of the order of a day and below.

Since we are interested in using $F_{TS}$ to construct very low FAR thresholds for our analysis, we need a good approximation of its distribution for extreme values.
This means in practice that we need to simulate and analyse a very long run of our experiment to produce enough statistics. In our simulations, we simulate between 25 and 100 years of background data.


Since the dataset modelling $F_{TS}$ was obtained through simulations, it means that it is completely dependent on the setup of the simulated experiment, namely the background rate $r_{bkg}$, the refresh time $t_r$, the window size $\omega$ and the definition of the test statistics $TS$.
If any of the simulation parameters are changed, the dataset needs to be recalculated. Out of the parameters listed above only one of them will not be specified by us during the real operation of the experiment, and that is the background rate $r_{bkg}$. The estimation of this parameter is discussed in detail in Sec.~\ref{sec:bkg_estimation}, but for now we can assume that it is known using a nominal value of $r_{bkg} = 0.18 \ \mathrm{cts} / \textrm{s}$. 
Finally, given a model of $F_{TS}$ and a false alarm interval $\tau_{false}$, expressed in seconds just like the refresh time, we can derive the corresponding threshold value $TS^*$ for the trigger:

\begin{equation}
    TS^* = F_{TS}^{-1} \left( \frac{t_r}{\tau_{false}} \right)
\end{equation}

\subsection{Trigger evaluation and comparison}
\label{sec:trigeval}
Given a specific setup of the experimental parameters $t_r$, $\omega$ and $TS$, we can assess the efficiency of the trigger by evaluating the rate of success when it comes to trigger activation in the presence of signal events. In order to test this, we will simulate experiments in which we inject events that follow a possible model distribution of neutrinos after a SN explosion, as shown in Fig.~\ref{fig:post_SN_dist}. In this we present results pertaining to a small selection of neutrino flare models, although the number of proposed models is abundant in literature and unknown in reality~\cite{Mirizzi:2015eza}.

\begin{figure}
    \centering
    \includegraphics[width=0.8\textwidth]{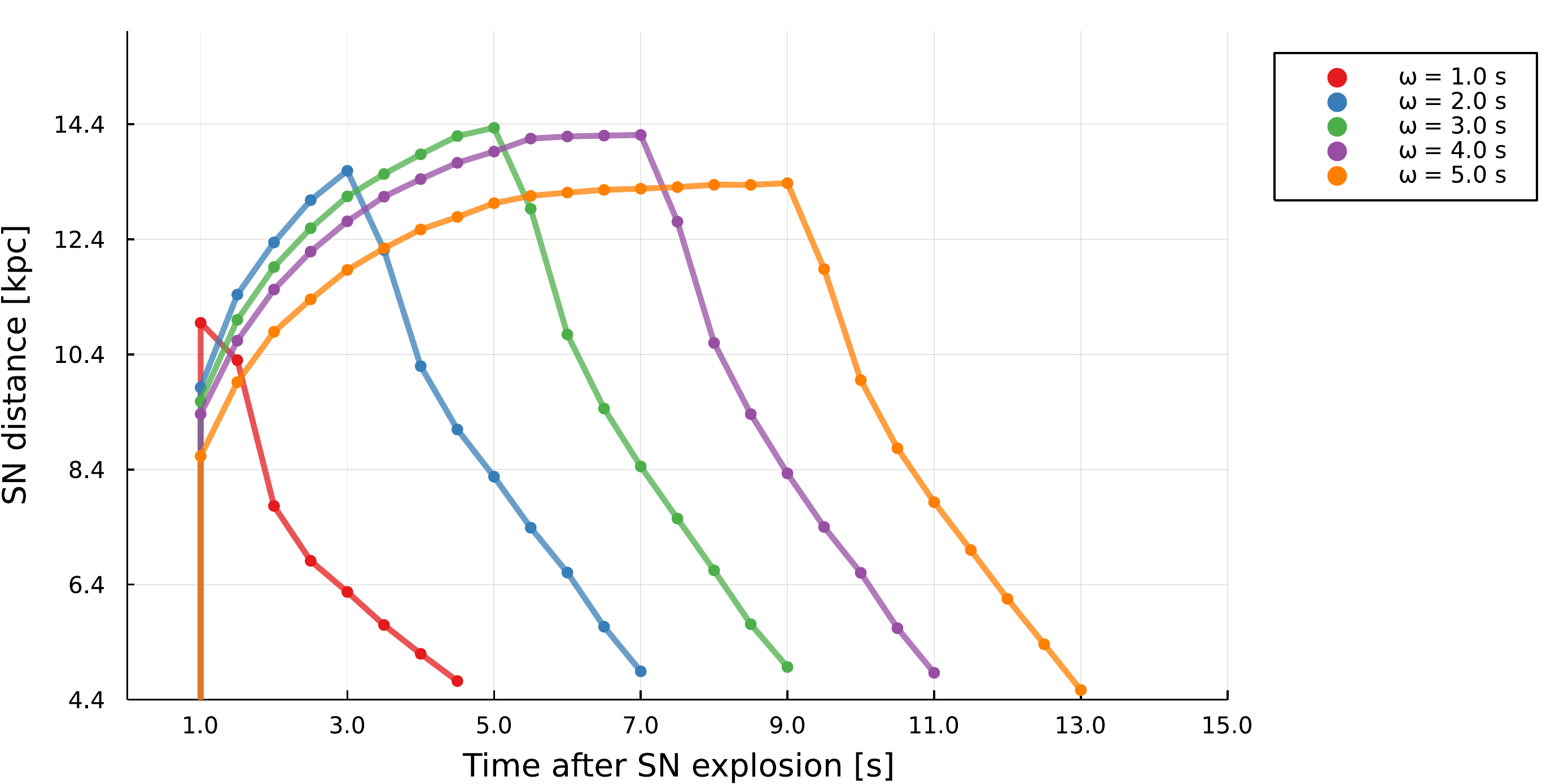}
    \caption{The 95\% quantile of successful SN detection distance, based on a FAR of 15 days, for different choices of the window size $w$. The refresh time is kept constant at 0.5\,s.}
    \label{fig:window_sizes}
\end{figure}

The number of neutrino events $\lambda_{sig}$ will be determined by the distance of the SN and after repeating these experiments for a large number of times, we can estimate the fraction of successful trigger activations (i.e. $1 - $type ii error). Fig.~\ref{fig:window_sizes} shows the maximum distance achieving a 95\% rate of success for different choices of the window size parameter $\omega$, as a function of the time after the explosion. The refresh time $t_r$ is kept at a constant 0.5\,s, an indicative value that matches the expected overall throughput of the raw data processing rate as discussed in Sec.~\ref{sec:time_resp}.

\begin{figure}
    \centering
    \includegraphics[width=0.8\textwidth]{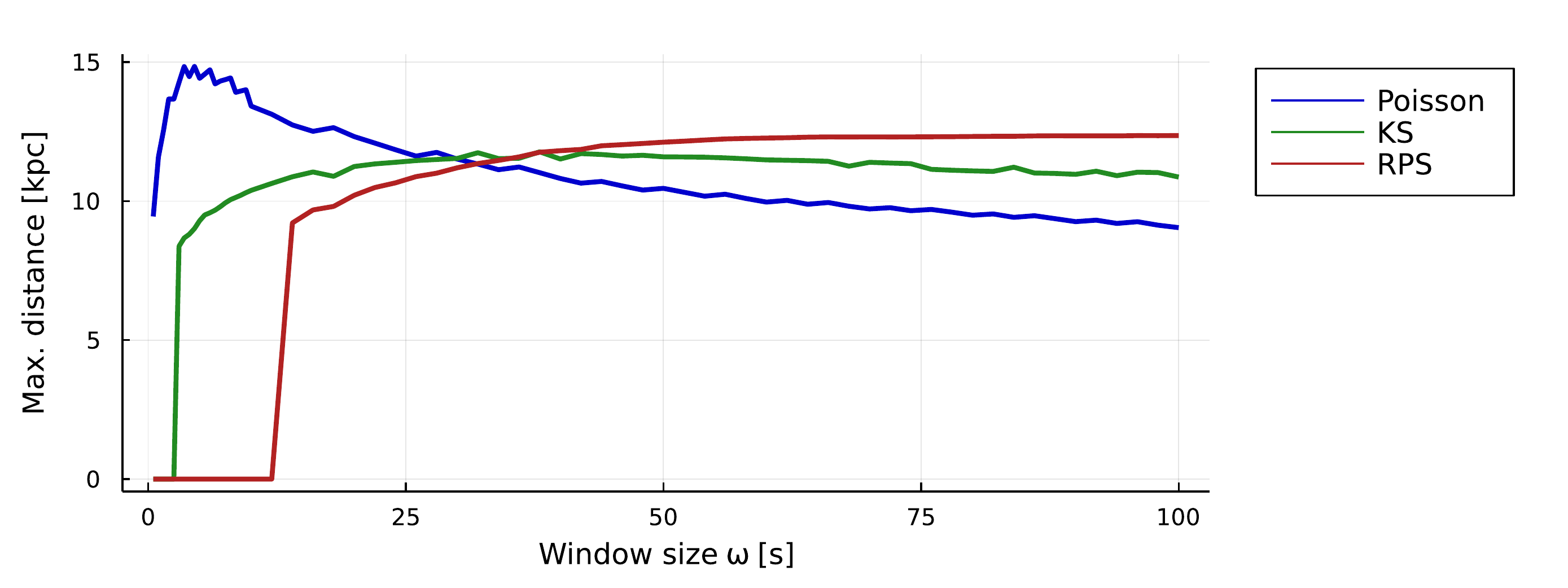}
    \caption{The 95\% quantile of successful detection distance for $r_{bkg} = 0.18$ cts/s, based on a FAR of 15 days, for different choices of the window size $\omega$. The refresh time is kept constant at 0.5\,s.}
    \label{fig:poisson_ks_rps}
\end{figure}

Examining Fig.~\ref{fig:window_sizes} we notice that around one second after the SN explosion the trigger starts to activate with a sharp turn on due to SN neutrinos. For very short windows, not all signal can be contained inside the window and the curve dies down rapidly again. For larger windows, further distances can be probed, since more of the signal can contribute to the statistic inside a window. For windows that are too large, however, more background events are being picked up that deteriorate the performance again. So there is an optimal window size, for the example in Fig.~\ref{fig:window_sizes} this lies at around 5~s.

\subsection{Non-parametric Tests as Alternatives to Poisson}

In the case of an optimal choice of window size and known background rate, the Poisson test will, in general, perform well. We have not found any alternative test outperforming an optimized Poisson test.
However, the Poisson test relies on the fact that the background rate is known or can be reliably estimated from data. This may not always be the case, or the background rate can even fluctuate. Furthermore, considering different signals of various time scales and shapes, or even a priori unknown transient signals, we want to explore alternatives to the Poisson test.

We investigate non-parametric tests as a viable alternative to, or used in combination with, Poisson. Non-parameteric tests are widely used in the scientific community for assessing the goodness-of-fit of a specific distribution given data. One of the most widely known tests is the Kolmogorov-Smirnov (KS)~\cite{Kolmogorov, Smirnov1948TableFE} test, which quantifies the agreement between the cumulative distribution and the empirical distribution. In our application, the null hypothesis is events only from background, resulting in a flat---i.e. uniform---distribution.

To compare and evaluate the performance of different tests, we consider a simulated experiment with a background rate of $r_{bkg} = 0.18 \ \mathrm{cts} /\mathrm{s}$, signal expectation of $\lambda_{sig} = 29.65 \, \mathrm{cts}$ at a reference distance of $10 \, \mathrm{kpc}$ and a refresh time $t_r = 0.5 \, \mathrm{s}$. We inspect the sensitivity of analysis windows of various sizes and perform a first screening by filtering for the SNe farthest detected at the 95\% success rate. To guarantee a fair comparison, the trigger thresholds for each test were evaluated in the same way as the one for the Poisson test, i.e., simulating an extended run of the experiment assuming a known background rate.
We further condense the information given by the success rate curve as in Fig.~\ref{fig:window_sizes} into a single number corresponding to the maximum distance that can be explored at the set success rate of 95\%. 
In our study, we have evaluated several such statistical tests, including Anderson-Darling (AD)~\cite{AD} and the Recursive Product of Spacings (RPS)~\cite{RPS}. 
The comparison of the resulting sensitivity of a selection of tests is shown in Fig.~\ref{fig:poisson_ks_rps}.
As we can see, for short analysis windows, the Poisson test outperforms the others, but as we increase the window size, the KS and RPS tests become more sensitive and yield better results. Looking at the furthest distance probed by each test for any given window size, the Poisson test appears to be the most sensitive, with the RPS test as a close second. Out of the test statistics we studied, the Poisson and the RPS tests excelled due to their sensitivity and in the following section we will show for non-optimal signal shape, window size, or background rate choices, RPS can in fact outperform Poisson.

\subsection{Application to prompt detection of SN neutrino emission}
\label{sec:post-SN}

\begin{figure}
    \centering
    \includegraphics[width=0.8\textwidth]{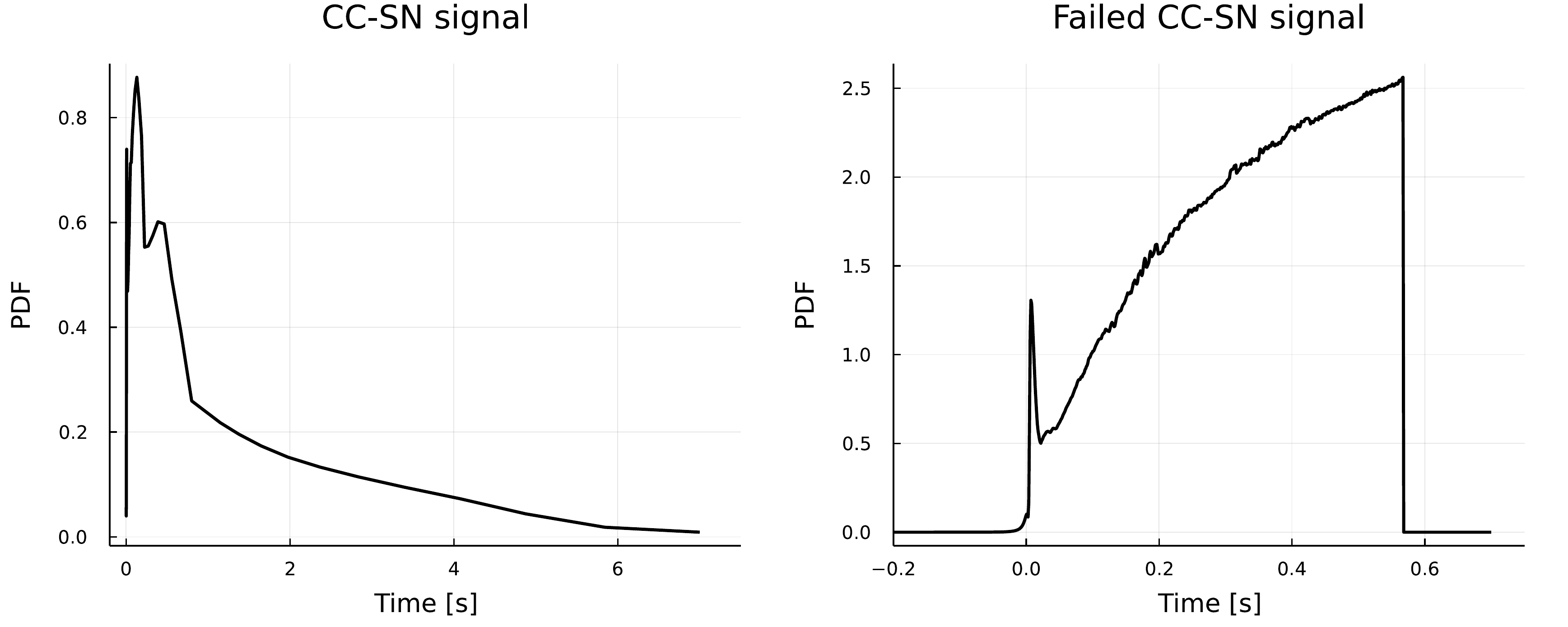}
    \caption{Normalized signal distributions for a core-collapse SN (left) and failed CC-SN (right), for progenitors stars with 27~$M_{\odot}$ and 40~$M_{\odot}$ respectively. These are 1D hydrodynamical simulations performed by the Garching group~\cite{garch} and named \texttt{s27\_ls220} and \texttt{s40\_s7b2c}.}
    \label{fig:SN_and_BH_signals}
\end{figure}

\begin{figure}
    \centering
    \includegraphics[width=1.0\textwidth]{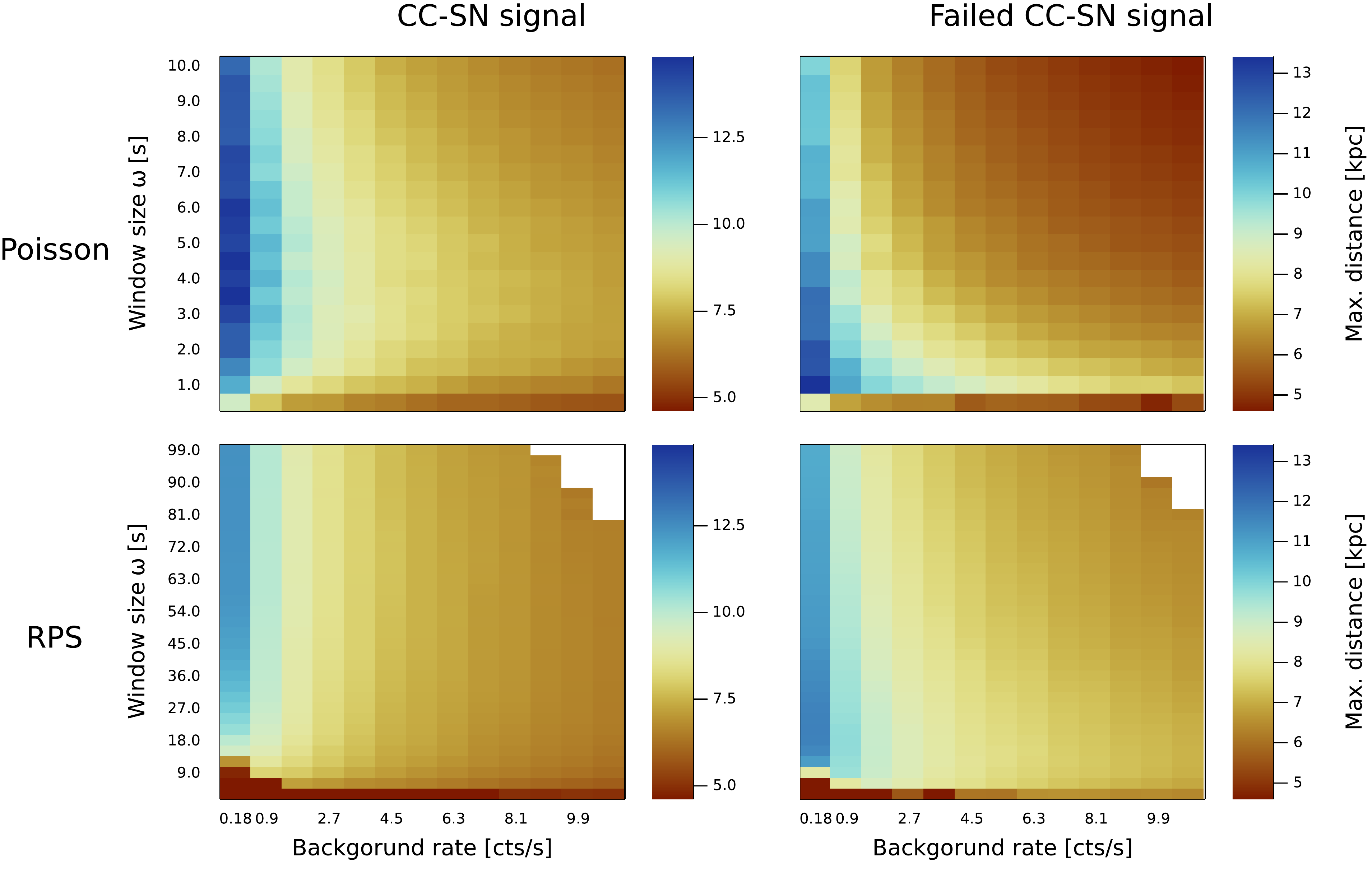}
    \caption{Maximum distance probed at 95\% success rate for different background rates and window sizes. The refresh time is kept constant at 0.5\,s. The white corners in the bottom row plots are due to the high total event expectation surpassing $10^3$, which is currently the upper limit when it comes to the parametrization of the RPS test.}
    \label{fig:window_sizes_background_rate}
\end{figure}

\begin{figure}
    \centering
    \includegraphics[width=0.8\textwidth]{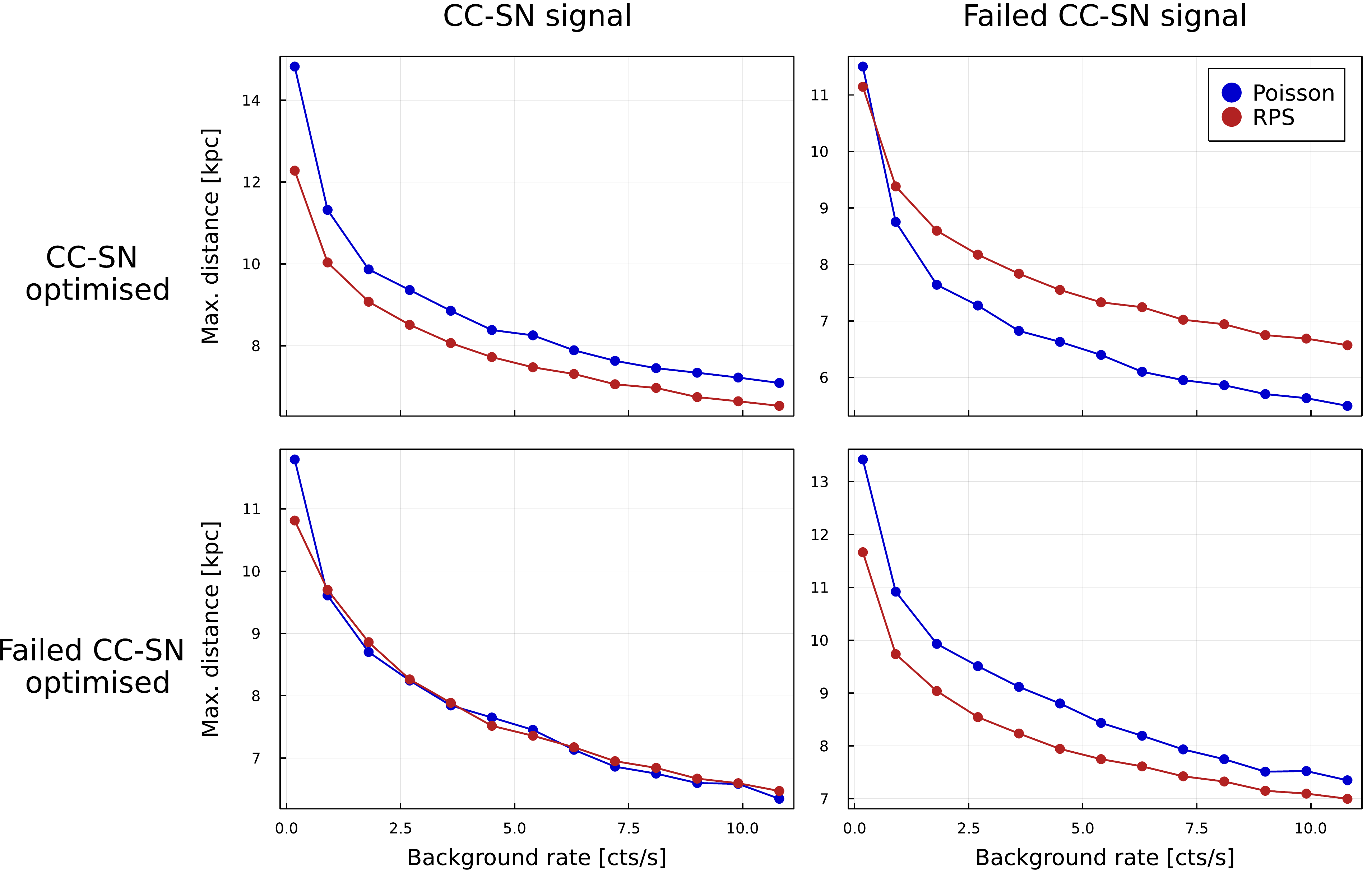}
    \caption{Maximum distance probed at a 95\% success rate as a function of time with respect to two sample signals, the Core-Collapse SN and the failed Core-Collapse SN, obtained using analysis windows optimised on each of the tested signals.}
    \label{fig:SN_and_BH_optimised_windows}
\end{figure}

Following up on the previous example of CC-SN neutrino detection, after analysing different window sizes $\omega$ and different test statistics, the best choice in terms of the furthest successfully detected SN signal out of the one shown in Fig.~\ref{fig:post_SN_dist} is a window of 5~s analysed with the Poisson test.
The signal distribution used in this example is just one possible signal that we would like to detect with our experiment. A very short list of possible SN models is available in \cite{EstrellaNueva}, where 30~models are presented with time distributions spanning from 0.5~s to 15.4~s. When selecting the correct analysis scheme, i.e., the window sizes and tests to use, we should also study the robustness of our choice against multiple models. As an example, we consider an alternative signal distribution, modelling a neutrino burst coming from a failed CC-SN event that results in the formation of a black hole, as depicted in the right panel of Fig.~\ref{fig:SN_and_BH_signals}. These models are 1D hydrodynamical simulations performed by the Garching group~\cite{garch}. They are the same adopted in~\cite{Pattavina:2020cqc} and named \texttt{LS220} and \texttt{failed-SN fast}, and they refer to progenitor stars with 27~$M_{\odot}$ and 40~$M_{\odot}$, respectively. In the latter case, the signal strength that is seen by the  detector is $\lambda_{sig} = 16.39 \, \mathrm{cts}$, weaker than the one induced by the CC-SN \texttt{LS220} model and this will result in shorter distances that can be probed by the detector. 
We repeat the same analysis previously described, namely, estimating the maximum distance at which we can reliably detect a neutrino burst with a 95\% success rate, while at the same time considering different values of the background rate, in order to account for possibly higher background levels in our experiment. We analyze both the CC-SN and the failed CC-SN signals shown in Fig.~\ref{fig:SN_and_BH_signals} using both the Poisson test and the RPS test, and the results of this study are shown in Fig.~\ref{fig:window_sizes_background_rate}.
Looking at these results, we notice that for both signals, as the background rate increases, the 95\% detection horizon starts to decrease. Given a fixed background rate, we notice that, while using the Poisson test, it is possible to achieve the furthest detection only for a select few analysis windows, while the horizon probed via the RPS test appears to be much more robust to changes of the window size, an effect that is particularly visible in the case of the failed CC-SN signal.
If we have detailed knowledge of the background affecting our experiment at any given time, and especially if we knew the time distributions of all the signals we might detect, then we could select the best combination of test statistic and window sizes. Looking at the results of Fig.~\ref{fig:window_sizes_background_rate} it would appear that the Poisson test is the most sensitive, provided that we have detailed knowledge of both the background and the signal. Although there are lists of different possible signals, in order to maximize the detection of all signals we might run a dedicate analysis for each proposed model. Such an approach would translate in running a multitude of parallel analysis streams, each with their own optimized window size for a given background rate. Our objective is to integrate our analysis with the SNEWS alert system; thus, we have to curtail the FAR of our final analysis. If we were to use independent analysis windows, then the FAR of each analysis stream would have to decrease proportionally to the total number of windows, which would discourage having too many of them. Additionally, such an approach may well not be the best suited one when we consider the sensitivity of our analysis to unknown signals, whose model was not considered during the development of the analysis scheme.
Furthermore, during the operation of our experiment, the background may realistically experience fluctuations in time, which could affect the sensitivity of the analysis windows. 
In order to limit the number of analysis windows we might need, and in order to retain good sensitivity against background fluctuations or with respect to unknown signals, we consider using few analysis streams, each with their own window size and using either the Poisson test or the RPS test. To gauge the potential and shortcomings of either test, we can test the sensitivity of the analyses optimized against a known signal with a known background against another signal distribution at different background levels. We present the results of this study in Fig.~\ref{fig:SN_and_BH_optimised_windows}. 
If we consider a CC-SN signal and a nominal background rate of 0.18 cts/s, we can select one of the best window size for each test, i.e., the window sizes that allow the detection of the furthest sources. Once we have selected one such window for each test, we increase the background rate and estimate the new horizon with 95\% trigger success. These results are reported in Fig.~\ref{fig:SN_and_BH_optimised_windows} (top-left) where we notice that both horizons are decreasing, as expected, while the Poisson one remains dominant.
If instead of the CC-SN signal we try to detect the failed CC-SN signal, while retaining the CC-SN-optimized windows, in Fig.~\ref{fig:SN_and_BH_optimised_windows} (top-right) we see that the horizon delivered by the RPS windows is roughly the same as the Poisson one for the nominal background rate and for higher backgrounds it becomes the better one, meaning that the RPS analysis proves to be more sensitive to narrower than expected signals compared to the Poisson analysis.
If we repeat the same analysis, this time optimizing with respect to the failed CC-SN signal and then testing against the CC-SN one, we notice that both analysis are equally sensitive to the latter signal.

These results, coupled with the overall picture presented in Fig.~\ref{fig:window_sizes_background_rate}, show that the RPS analysis can be more sensitive that the Poisson one when considering different signal distributions, as in a real-case scenario, especially when the detector operates in high background conditions.

\section{Triggers for pre-SN neutrinos}
\label{sec:trigger}

In the previous section, we described in detail our analysis scheme and the methods involved in estimating both the background neutrino rate and the threshold for a trigger system, using a combination of different test statistics. The examples we showed above presented the results for the detection of neutrinos produced shortly after the relevant event occurred, such as neutrino flares coming from SN explosions or BH formations. Here, we study the sensitivity of our experiment when it comes to the detection of neutrinos preceding a SN explosion. 
The time distribution of these precursor neutrinos is shown in Fig.~\ref{fig:recoiltimepreFlux}, where we clearly notice that this signal is rising exponentially as the SN explosion approaches.


When dealing with pre-SN neutrinos, the effective background rate we can estimate after the energy cuts is much smaller than the CC-SN case. For the purpose of the next example, we will consider a background rate $r_{bkg}$ = 0.018 \,cts/s, while the neutrinos rate coming from the signal distribution is $\lambda_{sig} \approx 154 \, \mathrm{cts}$ at the reference distance of $160 \, \mathrm{pc}$. Although the expected signal count is indeed quite high these precursor neutrinos we notice that they are mostly concentrated close to the explosion, given the highly exponential behavior of the distributions shown in Fig.~\ref{fig:recoiltimepreFlux}. 

\begin{figure}[htp]

\centering
\includegraphics[width=0.43373493975\textwidth]{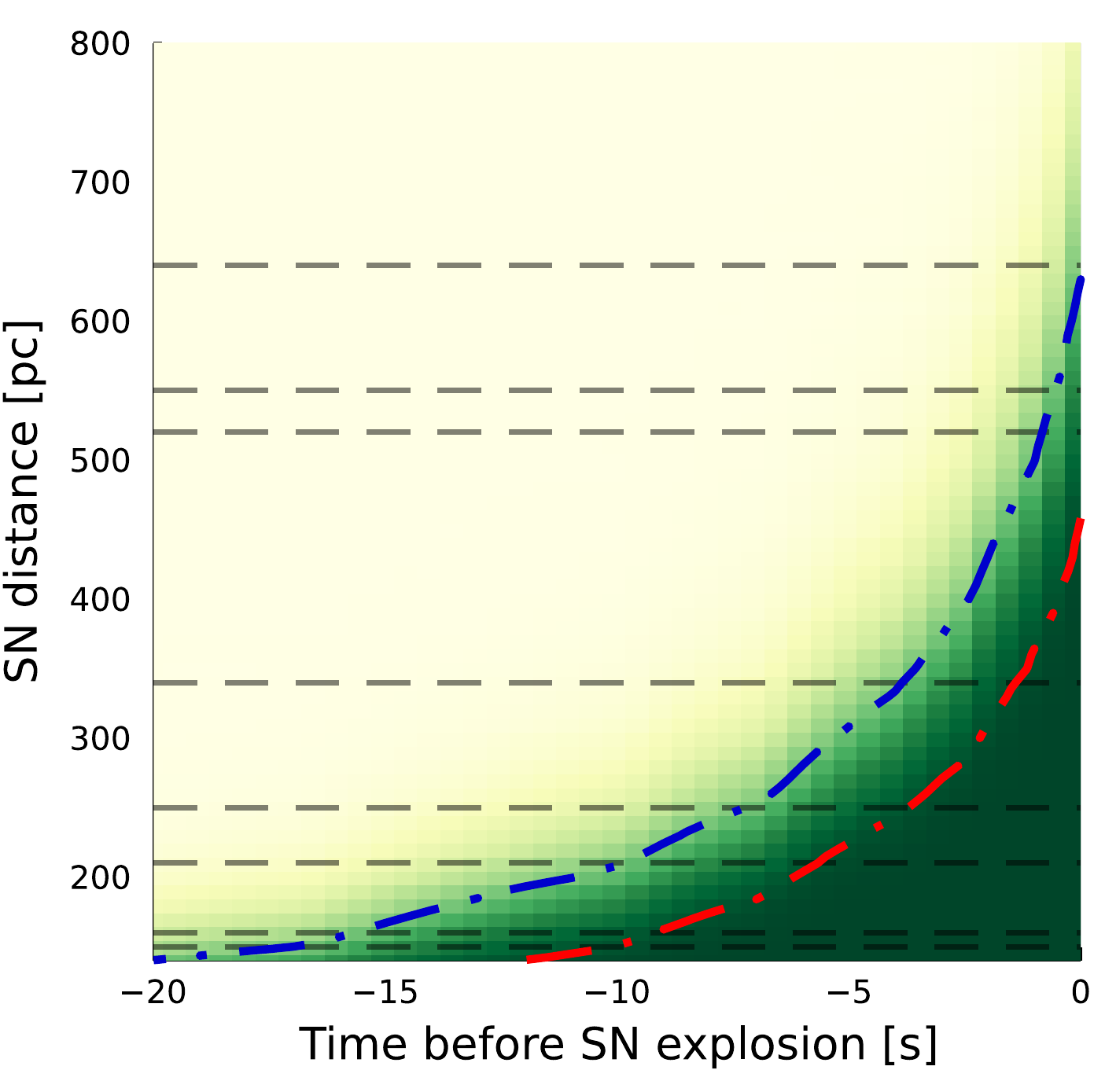}
\includegraphics[width=0.46626506024\textwidth]{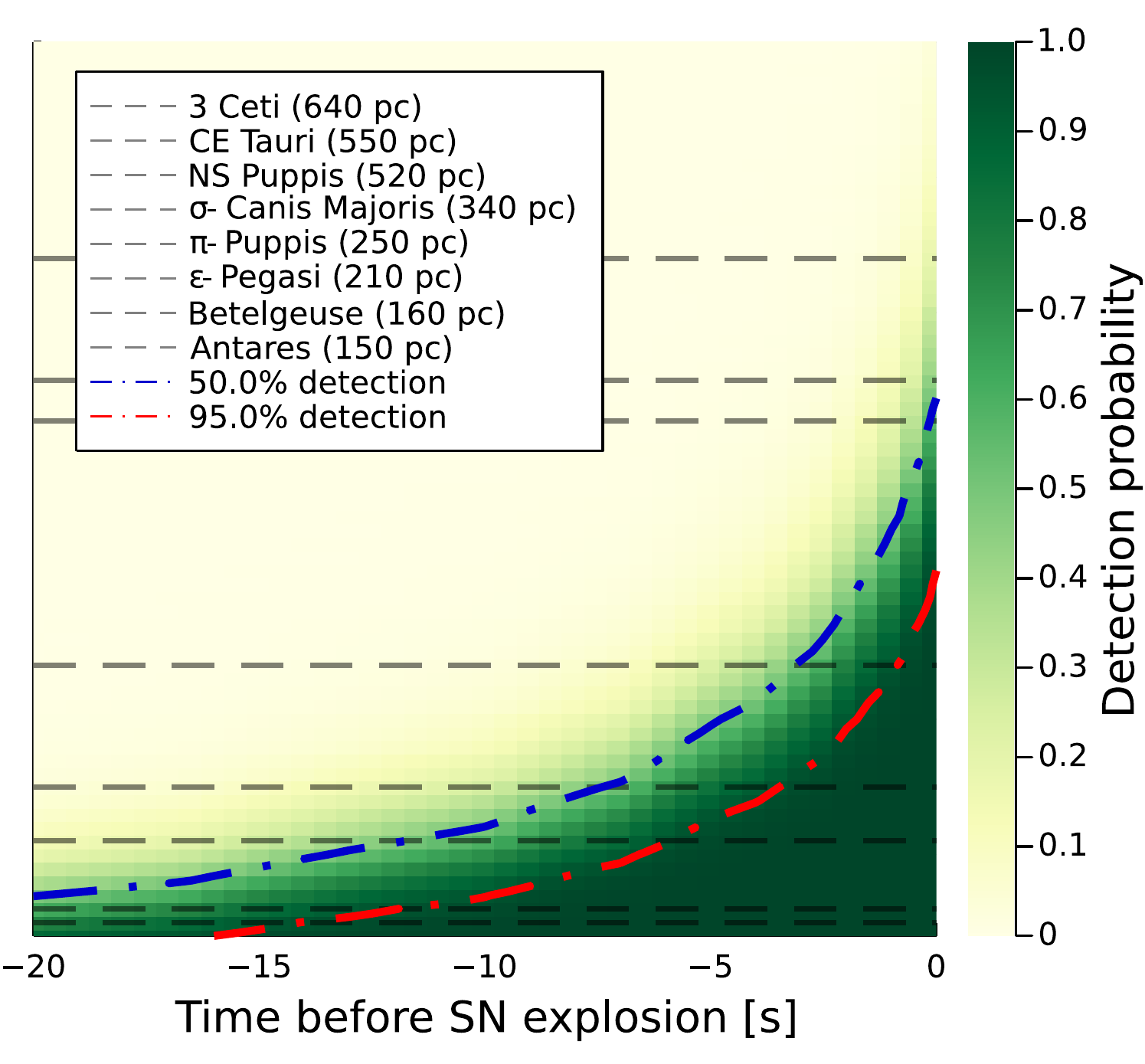}

\caption{Success rate of neutrinos detection preceding a SN explosion for three different window sizes: 15\,s (left); 70\,s (right).}
\label{fig:pre_SN_success_rate}

\end{figure}

As before, we can estimate the maximum distance of an SN explosion that would activate our trigger 95\% of times. Since we find ourselves in deep Poisson regime when considering the combined background and signal event counts, we present only the results reported in Fig.~\ref{fig:pre_SN_success_rate}, which shows a selection of the best analysis windows combined with the Poisson test.

For the pre-SN neutrinos, we have the option to optimize our analysis either towards the early detection of close sources, using short analysis windows ($\sim 15 \, \mathrm{s}$),  or towards the detection of the furthest achievable sources, using longer analysis windows ($\sim 70 \, \mathrm{s}$). These two cases are shown in the left and right panels of Fig.~\ref{fig:pre_SN_success_rate} respectively. Although the shorter window sizes are able to push the horizon slightly further, we notice that the horizon increases as we approach the SN explosion, making small gains less attractive.
Looking at Fig.~\ref{fig:pre_SN_success_rate}, we can state that our experiment is able to detect the precursor neutrinos up to $\sim 15 \, \mathrm{s}$ before the explosion with a success rate of 95\% at a distance of about $160 \, \mathrm{pc}$, approximately the distance at which Betelgeuse is located. Looking at the maximum probed distance, we notice that we are able to foretell a SN explosion circa $450 \, \mathrm{pc}$ away up to a few seconds before its occurrence.  



\section{Conclusions}
\label{sec:conclusions}
The next galactic SN will be an exceptional occasion to study the physics of these poorly understood fascinating phenomena, with exciting prospects in astronomy, astroparticle and particle physics. As of today, the technology to detect the gamma, gravitational waves, and neutrino signals emitted by such event is in place.

The coherent elastic neutrino-nucleus scattering is a very appealing detection channel, due to its large interaction cross-section and equal sensitivity to all neutrino flavours, for CC-SN neutrinos are supposed to populate roughly equally all the leptonic families.

In this work we have investigated the potential of low-threshold cryogenic particle detectors to detect the neutrino signal emitted prior the CC-SN event, and introduced a general framework to define the parameters of an online trigger for the  detection of pre-SN and CC-SN neutrinos in the context of an international networks, such as SNEWS.

We have also discussed the advantages of non-parametric test statistics for the application in a real-case scenario, where the details of the sought-for signal are unknown and not all the experimental conditions are fully under control. We have proved that RPS can outperform the widely utilized Poisson counting, when no specific prior is considered.

The application of this newly developed algorithm demonstrated the capability of the cm-sized RES-NOVA experiment to detect CC-SNe as far as 15~kpc with a 95\% success rate (and 20~kpc with 50\% success rate). This result underestimates previous calculations of RES-NOVA sensitivity to CC-SNe~\cite{Pattavina:2020cqc, RES-NOVA:2021gqp}. The discrepancy is due to the fact that in the present work, we investigate the live detection of CC-SN and failed CC-SN neutrinos without any knowledge of the onset time of the SN signal. This approach, which is more realistic, accounts for large background fluctuations that may mimic the SN signal while running the detector for long periods of time (type I and type II error). On the contrary, in the previous works based on maximum likelihood analyses, the sensitivity was evaluated knowing when there was a positive signal in the simulated data (type II error only). This is equivalent to consider the experiment ``on'' for only short time periods and, consequently, with limited background fluctuations. The investigation carried out in this work is an exhaustive study of what a real world experiment of such kind can achieve on its own.

At last, we showed that RES-NOVA can detect the feeble signal from pre-SN neutrinos at distances up to 450~pc and anticipate the CC-SN occurrence up to $\mathcal{O}(10~\textrm{s})$, when a very low detector energy threshold is achieved.

\section*{Acknowledgements}

This research was supported by the Deutsche Forschungsgemeinschaft (DFG, German Research Foundation) under Germany´s Excellence Strategy – EXC-2094 – 390783311 and the Sonderforschungsbereich (Collaborative Research Center) SFB1258 ‘Neutrinos and Dark Matter in Astro- and Particle Physics’.

\bibliographystyle{JHEP}
\bibliography{main}

\providecommand{\href}[2]{#2}\begingroup\raggedright\begin{thebibliography}{10}

\bibitem{Kamiokande1987}
K.~Hirata, T.~Kajita, M.~Koshiba, M.~Nakahata, Y.~Oyama, N.~Sato et~al.,
  \emph{Observation of a neutrino burst from the supernova sn1987a},
  \href{https://doi.org/10.1103/PhysRevLett.58.1490}{\emph{Phys. Rev. Lett.}
  {\bfseries 58} (1987) 1490}.

\bibitem{Review1987neutrino}
F.~Vissani and G.~Pagliaroli, \emph{Features of kamiokande-ii, imb, and baksan
  observations and their interpretation in a two-component model for the
  signal}, \href{https://doi.org/10.1134/S1063773709010010}{\emph{Astron.
  Lett.} {\bfseries 35} (2009) 1}.

\bibitem{IceCube:2018dnn}
{\scshape IceCube, Fermi-LAT, MAGIC, AGILE, ASAS-SN, HAWC, H.E.S.S., INTEGRAL,
  Kanata, Kiso, Kapteyn, Liverpool Telescope, Subaru, Swift NuSTAR, VERITAS,
  VLA/17B-403} collaboration, \emph{{Multimessenger observations of a flaring
  blazar coincident with high-energy neutrino IceCube-170922A}},
  \href{https://doi.org/10.1126/science.aat1378}{\emph{Science} {\bfseries 361}
  (2018) eaat1378} [\href{https://arxiv.org/abs/1807.08816}{{\ttfamily
  1807.08816}}].

\bibitem{TXS}
M.~Aartsen et~al., \emph{{Neutrino emission from the direction of the blazar
  TXS 0506+056 prior to the IceCube-170922A alert}},
  \href{https://doi.org/10.1126/science.aat2890}{\emph{Science} {\bfseries 361}
  (2018) 147} [\href{https://arxiv.org/abs/1807.08794}{{\ttfamily
  1807.08794}}].

\bibitem{Baade254}
W.~Baade and F.~Zwicky, \emph{On super-novae},
  \href{https://doi.org/10.1073/pnas.20.5.254}{\emph{Proceedings of the
  National Academy of Sciences} {\bfseries 20} (1934) 254}.

\bibitem{Mirizzi:2015eza}
A.~Mirizzi, I.~Tamborra, H.-T.~Janka, N.~Saviano, K.~Scholberg, R.~Bollig
  et~al., \emph{{Supernova Neutrinos: Production, Oscillations and Detection}},
  \href{https://doi.org/10.1393/ncr/i2016-10120-8}{\emph{Riv. Nuovo Cim.}
  {\bfseries 39} (2016) 1} [\href{https://arxiv.org/abs/1508.00785}{{\ttfamily
  1508.00785}}].

\bibitem{Nakamura_2016}
K.~Nakamura, S.~Horiuchi, M.~Tanaka, K.~Hayama, T.~Takiwaki and K.~Kotake,
  \emph{{Multimessenger signals of long-term core-collapse supernova
  simulations: synergetic observation strategies}},
  \href{https://doi.org/10.1093/mnras/stw1453}{\emph{Mon. Not. R. Astron. Soc.}
  {\bfseries 461} (2016) 3296}.

\bibitem{Odrzywolek:2003vn}
A.~Odrzywolek, M.~Misiaszek and M.~Kutschera, \emph{{Detection possibility of
  the pair-annihilation neutrinos from the neutrino-cooled pre-supernova
  star}},
  \href{https://doi.org/10.1016/j.astropartphys.2004.02.002}{\emph{Astropart.
  Phys.} {\bfseries 21} (2004) 303}
  [\href{https://arxiv.org/abs/astro-ph/0311012}{{\ttfamily
  astro-ph/0311012}}].

\bibitem{Asakura:2015bga}
{\scshape KamLAND} collaboration, \emph{{KamLAND Sensitivity to Neutrinos from
  Pre-Supernova Stars}},
  \href{https://doi.org/10.3847/0004-637X/818/1/91}{\emph{Astrophys. J.}
  {\bfseries 818} (2016) 91}
  [\href{https://arxiv.org/abs/1506.01175}{{\ttfamily 1506.01175}}].

\bibitem{Adams:2013ana}
S.M.~Adams, C.S.~Kochanek, J.F.~Beacom, M.R.~Vagins and K.Z.~Stanek,
  \emph{{Observing the Next Galactic Supernova}},
  \href{https://doi.org/10.1088/0004-637X/778/2/164}{\emph{Astrophys. J.}
  {\bfseries 778} (2013) 164}
  [\href{https://arxiv.org/abs/1306.0559}{{\ttfamily 1306.0559}}].

\bibitem{The:2006iu}
L.-S.~The, D.D.~Clayton, R.~Diehl, D.H.~Hartmann, A.F.~Iyudin, M.D.~Leising
  et~al., \emph{{Are ti-44 producing supernovae exceptional?}},
  \href{https://doi.org/10.1051/0004-6361:20054626}{\emph{Astron. Astrophys.}
  {\bfseries 450} (2006) 1037}
  [\href{https://arxiv.org/abs/astro-ph/0601039}{{\ttfamily
  astro-ph/0601039}}].

\bibitem{Firestone_2014}
R.B.~Firestone, \emph{{Observation of 23 Supernovae that exploded $<$300~pc
  from Earth during the past 300~kyr}},
  \href{https://doi.org/10.1088/0004-637x/789/1/29}{\emph{Astrophys. J.}
  {\bfseries 789} (2014) 29}.

\bibitem{Simpson:2019xwo}
{\scshape Super-Kamiokande} collaboration, \emph{{Sensitivity of
  Super-Kamiokande with Gadolinium to Low Energy Anti-neutrinos from
  Pre-supernova Emission}},
  \href{https://doi.org/10.3847/1538-4357/ab4883}{\emph{Astrophys. J.}
  {\bfseries 885} (2019) 133}
  [\href{https://arxiv.org/abs/1908.07551}{{\ttfamily 1908.07551}}].

\bibitem{Super-Kamiokande_PreSN}
{\scshape Super-Kamiokande} collaboration, \emph{{Sensitivity of
  Super-Kamiokande with Gadolinium to Low Energy Anti-neutrinos from
  Pre-supernova Emission}},
  \href{https://doi.org/10.3847/1538-4357/ab4883}{\emph{Astrophys. J.}
  {\bfseries 885} (2019) 133}
  [\href{https://arxiv.org/abs/1908.07551}{{\ttfamily 1908.07551}}].

\bibitem{Freedman:1973yd}
D.Z.~Freedman, \emph{{Coherent Neutrino Nucleus Scattering as a Probe of the
  Weak Neutral Current}},
  \href{https://doi.org/10.1103/PhysRevD.9.1389}{\emph{Phys. Rev. D} {\bfseries
  9} (1974) 1389}.

\bibitem{Raj:2019wpy}
N.~Raj, V.~Takhistov and S.J.~Witte, \emph{{Presupernova neutrinos in large
  dark matter direct detection experiments}},
  \href{https://doi.org/10.1103/PhysRevD.101.043008}{\emph{Phys. Rev. D}
  {\bfseries 101} (2020) 043008}
  [\href{https://arxiv.org/abs/1905.09283}{{\ttfamily 1905.09283}}].

\bibitem{Drukier:1983gj}
A.~Drukier and L.~Stodolsky, \emph{{Principles and Applications of a Neutral
  Current Detector for Neutrino Physics and Astronomy}},
  \href{https://doi.org/10.1103/PhysRevD.30.2295}{\emph{Phys. Rev.} {\bfseries
  D30} (1984) 2295}.

\bibitem{DM_cryo}
P.L.~Brink, \emph{Review of dark matter direct detection using cryogenic
  detectors}, \href{https://doi.org/10.1007/s10909-012-0517-7}{\emph{J. Low
  Temp. Phys.} {\bfseries 167} (2012) 1048}.

\bibitem{CUORE_Nature}
D.Q.~Adams, others and T.C.~Collaboration, \emph{Search for majorana neutrinos
  exploiting millikelvin cryogenics with cuore},
  \href{https://doi.org/10.1038/s41586-022-04497-4}{\emph{Nature} {\bfseries
  604} (2022) 53}.

\bibitem{Pattavina:2018nhk}
L.~Pattavina, M.~Laubenstein, S.S.~Nagorny, S.~Nisi, L.~Pagnanini, S.~Pirro
  et~al., \emph{{An innovative technique for the investigation of the 4-fold
  forbidden beta-decay of$^{50}$V}},
  \href{https://doi.org/10.1140/epja/i2018-12515-5}{\emph{Eur. Phys. J.}
  {\bfseries A54} (2018) 79}
  [\href{https://arxiv.org/abs/1801.03980}{{\ttfamily 1801.03980}}].

\bibitem{Casali:2013zzr}
N.~Casali et~al., \emph{{Discovery of the $^{151}$Eu $\alpha$ decay}},
  \href{https://doi.org/10.1088/0954-3899/41/7/075101}{\emph{J. Phys.}
  {\bfseries G41} (2014) 075101}
  [\href{https://arxiv.org/abs/1311.2834}{{\ttfamily 1311.2834}}].

\bibitem{Beeman:2011kv}
J.W.~Beeman et~al., \emph{{First measurement of the partial widths of
  $^{209}$Bi decay to the ground and to the first excited states}},
  \href{https://doi.org/10.1103/PhysRevLett.108.062501,
  10.1103/PhysRevLett.108.139903}{\emph{Phys. Rev. Lett.} {\bfseries 108}
  (2012) 062501} [\href{https://arxiv.org/abs/1110.3138}{{\ttfamily
  1110.3138}}].

\bibitem{Cardani:2013dia}
L.~Cardani et~al., \emph{{Development of a Li2MoO4 scintillating bolometer for
  low background physics}},
  \href{https://doi.org/10.1088/1748-0221/8/10/P10002}{\emph{JINST} {\bfseries
  8} (2013) P10002} [\href{https://arxiv.org/abs/1307.0134}{{\ttfamily
  1307.0134}}].

\bibitem{Beeman:2012wz}
J.W.~Beeman et~al., \emph{{New experimental limits on the alpha decays of lead
  isotopes}}, \href{https://doi.org/10.1140/epja/i2013-13050-7}{\emph{Eur.
  Phys. J.} {\bfseries A49} (2013) 50}
  [\href{https://arxiv.org/abs/1212.2422}{{\ttfamily 1212.2422}}].

\bibitem{Pattavina:2015jxe}
L.~Pattavina et~al., \emph{{Background Suppression in Massive TeO$_2$
  Bolometers with Neganov\textendash{}Luke Amplified Light Detectors}},
  \href{https://doi.org/10.1007/s10909-015-1404-9}{\emph{J. Low Temp. Phys.}
  {\bfseries 184} (2016) 286}
  [\href{https://arxiv.org/abs/1510.03266}{{\ttfamily 1510.03266}}].

\bibitem{Artusa:2016mat}
D.R.~Artusa et~al., \emph{{Enriched TeO$_2$ bolometers with active particle
  discrimination: towards the CUPID experiment}},
  \href{https://doi.org/10.1016/j.physletb.2017.02.011}{\emph{Phys. Lett. B}
  {\bfseries 767} (2017) 321}
  [\href{https://arxiv.org/abs/1610.03513}{{\ttfamily 1610.03513}}].

\bibitem{Kato:2017ehj}
C.~Kato, H.~Nagakura, S.~Furusawa, K.~Takahashi, H.~Umeda, T.~Yoshida et~al.,
  \emph{{Neutrino emissions in all flavors up to the pre-bounce of massive
  stars and the possibility of their detections}},
  \href{https://doi.org/10.3847/1538-4357/aa8b72}{\emph{Astrophys. J.}
  {\bfseries 848} (2017) 48}
  [\href{https://arxiv.org/abs/1704.05480}{{\ttfamily 1704.05480}}].

\bibitem{Patton:2015sqt}
K.M.~Patton, C.~Lunardini and R.J.~Farmer, \emph{{Presupernova neutrinos:
  realistic emissivities from stellar evolution}},
  \href{https://doi.org/10.3847/1538-4357/aa6ba8}{\emph{Astrophys. J.}
  {\bfseries 840} (2017) 2} [\href{https://arxiv.org/abs/1511.02820}{{\ttfamily
  1511.02820}}].

\bibitem{Odrzywolek:2009wa}
A.~Odrzywolek, \emph{{Nuclear statistical equilibrium neutrino spectrum}},
  \href{https://doi.org/10.1103/PhysRevC.80.045801}{\emph{Phys. Rev. C}
  {\bfseries 80} (2009) 045801}
  [\href{https://arxiv.org/abs/0903.2311}{{\ttfamily 0903.2311}}].

\bibitem{Joyce_2020}
M.~Joyce, S.-C.~Leung, L.~Moln{\'{a}}r, M.~Ireland, C.~Kobayashi and K.~Nomoto,
  \emph{Standing on the shoulders of giants: New mass and distance estimates
  for betelgeuse through combined evolutionary, asteroseismic, and hydrodynamic
  simulations with {MESA}},
  \href{https://doi.org/10.3847/1538-4357/abb8db}{\emph{Astrophys. J.}
  {\bfseries 902} (2020) 63}.

\bibitem{SNEWS:2020tbu}
{\scshape SNEWS} collaboration, \emph{{SNEWS 2.0: a next-generation supernova
  early warning system for multi-messenger astronomy}},
  \href{https://doi.org/10.1088/1367-2630/abde33}{\emph{New J. Phys.}
  {\bfseries 23} (2021) 031201}
  [\href{https://arxiv.org/abs/2011.00035}{{\ttfamily 2011.00035}}].

\bibitem{Pirro:2017ecr}
S.~Pirro and P.~Mauskopf, \emph{{Advances in Bolometer Technology for
  Fundamental Physics}},
  \href{https://doi.org/10.1146/annurev-nucl-101916-123130}{\emph{Ann. Rev.
  Nucl. Part. Sci.} {\bfseries 67} (2017) 161}.

\bibitem{Kim:2021wae}
Y.-H.~Kim, S.-J.~Lee and B.~Yang, \emph{{Superconducting detectors for rare
  event searches in experimental astroparticle physics}},
  \href{https://arxiv.org/abs/2111.08875}{{\ttfamily 2111.08875}}.

\bibitem{MUNSTER2017387}
A.~Muenster, S.~Schoenert and M.~Willers, \emph{Cryogenic detectors for dark
  matter search and neutrinoless double beta decay},
  \href{https://doi.org/10.1016/j.nima.2016.06.008}{\emph{Nucl. Instrum. Meth.
  A} {\bfseries 845} (2017) 387}.

\bibitem{Appec_DM}
J.~Billard, M.~Boulay, S.~Cebrian, L.~Covi, G.~Fiorillo, A.M.~Green et~al.,
  \emph{Direct detection of dark matter - appec committee report},
  {\emph{Reports on Progress in Physics} (2022) }.

\bibitem{Abdelhameed:2019hmk}
{\scshape CRESST} collaboration, \emph{{First results from the CRESST-III
  low-mass dark matter program}},
  \href{https://doi.org/10.1103/PhysRevD.100.102002}{\emph{Phys. Rev.}
  {\bfseries D100} (2019) 102002}
  [\href{https://arxiv.org/abs/1904.00498}{{\ttfamily 1904.00498}}].

\bibitem{CRESST:2019mle}
{\scshape CRESST} collaboration, \emph{{First results on sub-GeV spin-dependent
  dark matter interactions with$^{7}$Li}},
  \href{https://doi.org/10.1140/epjc/s10052-019-7126-4}{\emph{Eur. Phys. J. C}
  {\bfseries 79} (2019) 630}
  [\href{https://arxiv.org/abs/1902.07587}{{\ttfamily 1902.07587}}].

\bibitem{SuperCDMS:2022kse}
{\scshape SuperCDMS} collaboration, \emph{{A Strategy for Low-Mass Dark Matter
  Searches with Cryogenic Detectors in the SuperCDMS SNOLAB Facility}},  in
  \emph{{2022 Snowmass Summer Study}}, 3, 2022
  [\href{https://arxiv.org/abs/2203.08463}{{\ttfamily 2203.08463}}].

\bibitem{Armengaud:2019kfj}
{\scshape EDELWEISS} collaboration, \emph{{Searching for low-mass dark matter
  particles with a massive Ge bolometer operated above-ground}},
  \href{https://doi.org/10.1103/PhysRevD.99.082003}{\emph{Phys. Rev.}
  {\bfseries D99} (2019) 082003}
  [\href{https://arxiv.org/abs/1901.03588}{{\ttfamily 1901.03588}}].

\bibitem{Abdelhameed:2019oxl}
{\scshape CRESST} collaboration, \emph{{Geant4-based electromagnetic background
  model for the CRESST dark matter experiment}},
  \href{https://doi.org/10.1140/epjc/s10052-019-7385-0,
  10.1140/epjc/s10052-019-7504-y}{\emph{Eur. Phys. J.} {\bfseries C79} (2019)
  881} [\href{https://arxiv.org/abs/1908.06755}{{\ttfamily 1908.06755}}].

\bibitem{Pattavina:2020cqc}
L.~Pattavina, N.~Ferreiro~Iachellini and I.~Tamborra, \emph{{Neutrino
  observatory based on archaeological lead}},
  \href{https://doi.org/10.1103/PhysRevD.102.063001}{\emph{Phys. Rev. D}
  {\bfseries 102} (2020) 063001}
  [\href{https://arxiv.org/abs/2004.06936}{{\ttfamily 2004.06936}}].

\bibitem{Pattavina:2019pxw}
L.~Pattavina, J.W.~Beeman, M.~Clemenza, O.~Cremonesi, E.~Fiorini, L.~Pagnanini
  et~al., \emph{{Radiopurity of an archeological Roman Lead cryogenic
  detector}}, \href{https://doi.org/10.1140/epja/i2019-12809-0}{\emph{Eur.
  Phys. J.} {\bfseries A55} (2019) 127}
  [\href{https://arxiv.org/abs/1904.04040}{{\ttfamily 1904.04040}}].

\bibitem{RES-NOVA:2021gqp}
{\scshape RES-NOVA} collaboration, \emph{{RES-NOVA sensitivity to core-collapse
  and failed core-collapse supernova neutrinos}},
  \href{https://doi.org/10.1088/1475-7516/2021/10/064}{\emph{JCAP} {\bfseries
  10} (2021) 064} [\href{https://arxiv.org/abs/2103.08672}{{\ttfamily
  2103.08672}}].

\bibitem{RES-NOVAgroupofinterest:2022pvc}
{\scshape RES-NOVA} collaboration, \emph{{Radiopurity of a kg-scale PbWO$_4$
  cryogenic detector produced from archaeological Pb for the RES-NOVA
  experiment}},  \href{https://arxiv.org/abs/2203.07441}{{\ttfamily
  2203.07441}}.

\bibitem{Iachellini:2021rmh}
N.F.~Iachellini et~al., \emph{{Operation of an archaeological lead PbWO$_4$
  crystal to search for neutrinos from astrophysical sources with a Transition
  Edge Sensor}},  in \emph{{19th International Workshop on Low Temperature
  Detectors}}, 11, 2021 [\href{https://arxiv.org/abs/2111.07638}{{\ttfamily
  2111.07638}}].

\bibitem{CRESST:2015txj}
{\scshape CRESST} collaboration, \emph{{Results on light dark matter particles
  with a low-threshold CRESST-II detector}},
  \href{https://doi.org/10.1140/epjc/s10052-016-3877-3}{\emph{Eur. Phys. J. C}
  {\bfseries 76} (2016) 25} [\href{https://arxiv.org/abs/1509.01515}{{\ttfamily
  1509.01515}}].

\bibitem{Reindl:2016yhs}
F.~Reindl, \emph{{Exploring Light Dark Matter With CRESST-II Low-Threshold
  Detectors}}, Ph.D. thesis, Munich, Tech. U., 2016.

\bibitem{Franz}
F.~Pr{\"o}bst et~al., \emph{Model for cryogenic particle detectors with
  superconducting phase transition thermometers},
  \href{https://doi.org/10.1007/BF00753837}{\emph{J. Low Temp. Phys.}
  {\bfseries 100} (1995) 69}.

\bibitem{CRESST2}
G.~Angloher et~al., \emph{Results from 730 kg days of the cresst-ii dark matter
  search}, \href{https://doi.org/10.1140/epjc/s10052-012-1971-8}{\emph{Eur.
  Phys. J. C} {\bfseries 72} (2012) 1971}.

\bibitem{FerreiroIachellini:2019obk}
N.~Ferreiro~Iachellini, \emph{{Increasing the sensitivity to low mass dark
  matter in CRESST-III with a new DAQ and signal processing}}, Ph.D. thesis,
  Munich U., 2019.
\newblock 10.5282/edoc.23762.

\bibitem{Gibbs}
J.W.~Gibbs, \emph{Fourier's series},
  \href{https://doi.org/10.1038/059200b0}{\emph{Nature} {\bfseries 59} (1898)
  200}.

\bibitem{Agafonova:2007hn}
N.Y.~Agafonova et~al., \emph{{On-line recognition of supernova neutrino bursts
  in the LVD detector}},
  \href{https://doi.org/10.1016/j.astropartphys.2007.09.005}{\emph{Astropart.
  Phys.} {\bfseries 28} (2008) 516}
  [\href{https://arxiv.org/abs/0710.0259}{{\ttfamily 0710.0259}}].

\bibitem{Lamoureux:2021bat}
M.~Lamoureux, \emph{{Identification of neutrino bursts associated to supernovae
  with real-time test statistic (RTS2) method}},
  \href{https://doi.org/10.1051/0004-6361/202141305}{\emph{Astron. Astrophys.}
  {\bfseries 654} (2021) A95}
  [\href{https://arxiv.org/abs/2103.09733}{{\ttfamily 2103.09733}}].

\bibitem{Kolmogorov}
A.~Kolmogorov-Smirnov et~al., \emph{Sulla determinazione empirica di una legge
  di distribuzione}, {\emph{G. Ist. Ital.} {\bfseries 4} (1933) 83–91}.

\bibitem{Smirnov1948TableFE}
N.~Smirnov, \emph{Table for estimating the goodness of fit of empirical
  distributions}, {\emph{Ann. Math. Stat.} {\bfseries 19} (1948) 279}.

\bibitem{AD}
T.W.~Anderson and D.A.~Darling, \emph{A test of goodness of fit},
  \href{https://doi.org/10.1080/01621459.1954.10501232}{\emph{J. Am. Stat.
  Assoc.} {\bfseries 50} (1954) 765}.

\bibitem{RPS}
P.~Eller and L.~Shtembari, \emph{{A goodness-of-fit test based on a recursive
  product of spacings}},  \href{https://arxiv.org/abs/2111.02252}{{\ttfamily
  2111.02252}}.

\bibitem{EstrellaNueva}
O.I.~Gonzalez-Reina et~al., \emph{Estrellanueva: an open-source software to
  study the interactions and detection of neutrinos emitted by supernovae},
  2022.
\newblock 10.5281/zenodo.6354850.

\bibitem{garch}
{MPA Supernova Archive, \url{https://wwwmpa.mpa-garching.mpg.de/ccsnarchive}}.
  \url{https://wwwmpa.mpa-garching.mpg.de/ccsnarchive/data/}.

\end{thebibliography}\endgroup

\end{document}